\documentclass[fleqn,usenatbib,single-column]{mnras}
\usepackage{newtxtext,newtxmath}
\usepackage{orcidlink}

\usepackage[T1]{fontenc}

\DeclareRobustCommand{\VAN}[3]{#2}
\let\VANthebibliography\thebibliography
\def\thebibliography{\DeclareRobustCommand{\VAN}[3]{##3}\VANthebibliography}

\usepackage{graphicx}	
\usepackage{amsmath}	

\newcommand{\Cloudy}{\textsc{Cloudy}}

\font\manual=manfnt at 7pt \def\dbend{\hbox{\raise0.9ex\hbox{\manual\char127\hspace{0.6em}}}}

\newcounter{INTERNALionstage}

\newcommand{\TODO}[1]{\textcolor{red}{{\textbf{\boldmath [#1]}}}}

\def\gtsim{\mathrel{\hbox{\rlap{\hbox{\lower4pt\hbox{$\sim$}}}\hbox{$>$}}}}
\def\lesssim{\mathrel{\hbox{\rlap{\hbox{\lower4pt\hbox{$\sim$}}}\hbox{$<$}}}}

\def\ga{{\rm\thinspace gauss}}

\def\la{\mbox{{\rm L}$\alpha$}}

\def\hii{\mbox{{\rm H~{\sc ii}}}}

\def\h0{\mbox{{\rm H}$^0$}}
\DeclareMathAlphabet{\vib}{OML}{cmm}{m}{it}

\title[
JWST/NIRCam selected galaxies at $z\ >$ 11]{Detection of New Galaxy Candidates at $z\ >$ 11 in the JADES Field Using JWST NIRCam 
}

\author[Chakraborty et al.]{
Priyanka Chakraborty$^{1, 2}$\,\orcidlink{0000-0002-4469-2518},\thanks{E-mail: priyanka.chakraborty@cfa.harvard.edu}
Arnab Sarkar$^{2,3}$\,\orcidlink{0000-0002-5222-1337},
Scott Wolk$^{1}$\,\orcidlink{0000-0002-0826-9261},
Benjamin Schneider$^{3}$\,\orcidlink{0000-0003-4876-7756},
Nancy Brickhouse$^{1}$\,\orcidlink{0000-0002-8704-4473},
\newauthor
Kenneth Lanzetta$^{4}$,
Adam Foster$^{1}$\,\orcidlink{0000-0003-3462-8886},
Randall Smith$^{1}$\,\orcidlink{0000-0003-4284-4167}
\\\\
$^{1}$Center for Astrophysics $\vert$ Harvard \& Smithsonian, Cambridge, MA 02138, USA\\
$^{2}$Department of Physics, University of Arkansas, 825 West Dickson Street, Fayetteville, AR 72701, USA\\
$^{3}$Kavli Institute for Astrophysics and Space Research,
Massachusetts Institute of Technology, 70 Vassar St, Cambridge, MA 02139\\
$^{4}$ Stony Brook University, Stony Brook, NY 11794-3800\\
}

\date{Accepted XXX. Received YYY; in original form ZZZ}

\pubyear{2024}

\begin{document}
\label{firstpage}
\pagerange{\pageref{firstpage}--\pageref{lastpage}}
\maketitle

\begin{abstract}
We report the detection of { six}
new galaxy
candidates with 
{ photometric}
redshifts $z$ $>$ 11
within the JWST Advanced 
Deep Extragalactic Survey (JADES) 
GOODS-S
and GOODS-N fields.
These new candidates are detected 
through meticulous analysis of
NIRCam photometry in eight filters
spanning a wavelength range of 
0.8-5.0 $\mu$m. 
Photometric redshifts of these
galaxy candidates are independently
measured utilizing
spectral energy distribution (SED) fitting techniques using 
\texttt{EAZY} and 
\texttt{BAGPIPES} codes, followed by visual
scrutiny. 
One of these galaxy candidates 
is located in GOODS-S field, while
the remaining five galaxies
are located in GOODS-N
field. Our analysis reveals that the
stellar masses of these galaxies
typically
range from log $M_{\ast}$/$M_{\odot}$
= 7.75--8.75. 
Futhermore, these galaxies are
typically young with their 
mass-weighted ages
spanning from 80 to 240 Myr. 
Their
specific star formation rates (sSFR)
, quantified 
as $\log (\text{sSFR}/\text{Gyr}$), 
are measured to vary between 
$\sim 0.95$ to 1.46. 
These new galaxy candidates
offer a robust
sample for probing the physical properties
of galaxies within the first
few hundred Myr of the 
history of the Universe. We also analyze the relationship between star formation rate (SFR) and stellar mass ($M_\ast$) within our sample.
Continued investigation through 
spectroscopic analysis using 
JWST/NIRSpec is needed to 
spectroscopically confirm these 
high-redshift galaxy candidates 
and investigate
further into their physical properties.  We plan to follow up on these candidates with future NIRSpec observations.
\end{abstract}




\section{Introduction}

The galaxies that emerged within the
first few hundred million years
after recombination are believed to
have played a crucial role in shaping
the early Universe by hosting its 
first stars. 
This first generation of stars
initiated the process of ionizing 
neutral hydrogen throughout the cosmos. 
While reionization is generally
understood to have occurred within the 
first billion years following 
the Big Bang, the precise mechanisms
and the specific types of galaxies 
responsible for this phenomenon remain largely elusive and subject to 
ongoing debate
\citep[e.g.,][]{2001PhR...349..125B,2004ARA&A..42...79B,2006ARA&A..44..415F,2011ARA&A..49..373B,2016ARA&A..54..761S,2020ARA&A..58..617O}.
Thus, studying early galaxies is crucial
for understanding their formation and
evolution over billions of years,
revealing insights into the physical
processes that shaped them
and their role in re-ionizing the Universe.

Over the past decade, deep surveys conducted
by the Hubble Space Telescope (HST) 
discovered thousands of galaxies, 
significantly improving our understanding
of the properties and demographics of 
galaxies up to z $\sim$ 11  
\citep{2013ApJ...763L...7E, 2015ApJ...799...12I, 2015ApJ...810...71F, 2016MNRAS.459.3812M, 2019MNRAS.486.3805B, 2021AJ....162...47B}. 
These surveys included the
Hubble Ultra-Deep Field (HUDF, 
\citealt{2006AJ....132.1729B}),
HUDF09 \citep{2011ApJ...737...90B},
HUDF12 \citep{2013ApJ...763L...7E},
the HST Great Observatories Origins Deep 
Survey (GOODS, \citealt{2004ApJ...600L..93G}),
and many others 
\citep[e.g.,][]{2007ApJS..172....1S,2019ApJ...884...85C}.
However, the limited near-infrared (NIR) 
coverage of the HST, along with
its moderate light-collecting capacity,
limited the exploration of the evolution
of galaxies at earlier ($z$ $>$ 11) epochs.

The  method for selecting 
high-redshift galaxies has primarily been
driven by photometry 
\citep{2013ApJ...768..196S, 2015ApJ...803...34B}. 
This technique exploits the absorption of 
ultraviolet light by neutral hydrogen within, 
around, and between distant galaxies. This 
absorption produces a distinctive spectral feature 
known as the ``{ Lyman-$\alpha$ break}," observable at 912  
\AA\ in the spectral energy distribution (SED) of 
the galaxy
\citep{2019PASP..131g4101S}.
The redshifts of these galaxies are 
determined by fitting their observed
SEDs to 
either simulated or observed galaxy SEDs 
with known physical properties. 
This method incorporates more data
than pure color selection alone 
\citep{2000A&A...363..476B}. 
To effectively apply this approach,
accurate template SEDs covering the 
entire color space of photometry are 
necessary, accounting for factors 
such as dust extinction and 
intergalactic medium (IGM) absorption. 
However, this procedure becomes 
less certain for high-redshift
galaxies due to the scarcity of 
ultraviolet (UV) and optical SEDs for
early galaxies
\citep[e.g.,][]{2016ARA&A..54..761S}.

The commissioning of the James Webb Space Telescope (JWST) \citep{2023PASP..135d8001R}  has opened a new window into the discovery and study of galaxies during the earliest stages of the universe, made possible by its improved sensitivity, 7$\times$ light-collecting area, and the broad imaging field of  NIRCam \citep{2005SPIE.5904....1R}. In the early months of operation, several $z$ $>$ 11 galaxy candidates were identified \citep{2022ApJ...940L..14N, 2022ApJ...940L..55F, 2023ApJS..265....5H, 2023MNRAS.518.6011D, 2023MNRAS.519.1201A} as part of JWST's Early Release Observations (ERO; \citet{2022ApJ...936L..14P}), the Early Release Science (ERS) programs CEERS \citep{2023ApJ...946L..12B} and Through the Looking GLASS (GLASS-JWST) programs \citep{2022ApJ...935..110T}. 
Many of the galaxy candidates identified in these early observations exceeded the previous distance record established by HST studies \citep{2016ApJ...819..129O}. The discovery of early galaxy candidates has been further advanced by observational campaigns such as the JWST Advanced Deep Extragalactic Survey
\citep[JADES,][]{2023arXiv230602465E}
, the Next Generation Deep Extragalactic Exploratory Public \citep[NGDEEP,][]{2023ApJ...952L...7A}  survey, and the Prime Extragalactic Areas for Reionization and Lensing Science (PEARLS) project \citep{2023AJ....165...13W, 2023MNRAS.525.1353J}, which have utilized photometric techniques to identify z $\sim$ 9-16 galaxy candidates.


JADES is one of the most extensive initiatives in JWST's first year of observation, dedicating 32 days of telescope time to the discovery and characterization of the most distant galaxies \citep{2023PASP..135b8001R}. 
The survey encompasses two deep fields, GOODS-South and GOODS-North, which have extensive legacy datasets provided by the Great Observatories Origins Deep Survey \citep[GOODS;][]{2003mglh.conf..324D, 2004ApJ...600L..93G}, and utilizes three instruments: NIRCam for imaging, NIRSpec for spectroscopy, and MIRI with an imaging and spectroscopy. JADES operates at two levels of survey depth: "Deep" and "Medium." The Deep survey is designed for in-depth characterization of fainter galaxies or detailed analysis of individual sources at higher signal-to-noise ratios, whereas the Medium survey facilitates a statistical analysis of the high-redshift galaxy population.
{ Recently, \citet{2024ApJ...964...71H} presents
a impressive catalog of
717 candidate galaxies at redshifts $z>8$, identified using deep imaging from JWST JADES program along with complementary data from FRESCO, JEMS, and HST. 
The candidates, selected through photometric redshifts, include many newly discovered sources—some potentially at redshifts up to $z\sim18$.}

In this paper, we report the detection
of new $z$ $>$ 11 galaxy candidates using NIRCam observations taken across eight filters from the first year of the JADES survey in the GOODS-S and GOODS-N fields. 
{ The targets analyzed in this study are also included in the 
catalog released by JADES team, although with lower redshift estimates. Our analysis provides independent redshift measurements for these sources, based on refined photometry and an alternative modeling approach, offering a complementary perspective on their physical properties.}
We use two distinct codes \texttt{EAZY} \citep{2008ApJ...686.1503B} and \texttt{BAGPIPES} \citep{2018MNRAS.480.4379C} for determining the redshifts of the candidates and their physical properties using SED fitting techniques to the photometric data.
The format of the paper is as follows. In Section \ref{data}, we provide a detailed description of the observation and the photometric data reduction process, including photometric redshift estimation and dropout selection methods. 
In section \ref{selection}, we detail the methodology for selecting galaxies at redshifts $z$ $>$ 11. 
In Section \ref{results}, we present the main results from our galaxy sample, including detailed descriptions, sizes, properties derived from SED fitting, and an analysis of systematic uncertainties and potential sources of contamination. In Section \ref{con}, we summarize our findings.

Throughout this paper, we adhere to the AB magnitude system  \citep{1983ApJ...266..713O} and cosmological parameters reported by  \citet{2020A&A...641A...6P}: the Hubble constant \( H_0 = 67.4 \, \text{km s}^{-1} \text{Mpc}^{-1} \), matter density parameter \( \Omega_M = 0.315 \), and dark energy density \( \Omega_\Lambda = 0.685 \).

\begin{figure*}
\includegraphics[width=1\textwidth]{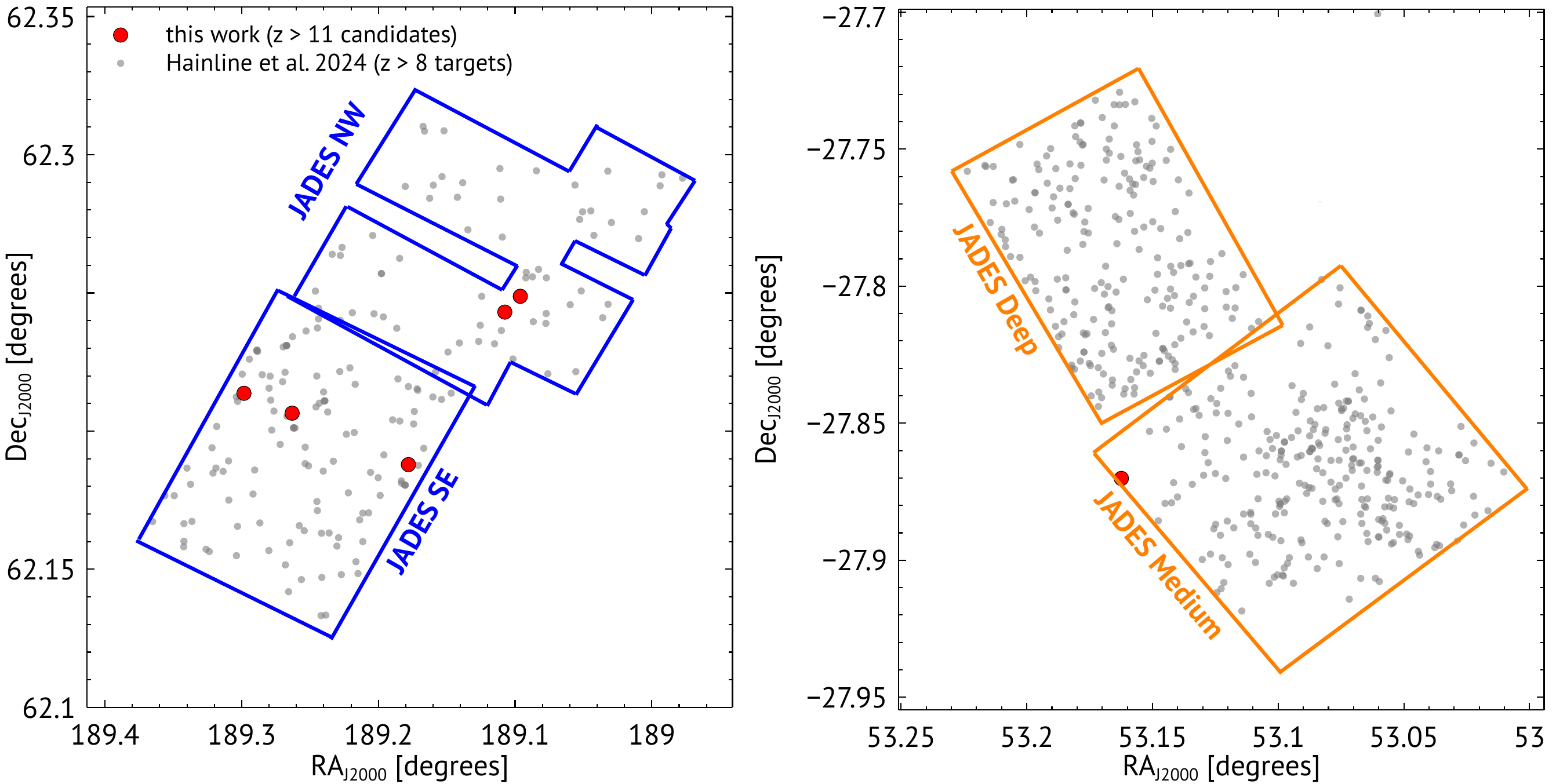}
    \caption{Left: Map displaying the GOODS-N field. The red data points represent our newly identified targets with  z$>$11 overlaid on the grey data points showing candidates with z$>$8 listed in \citet{2024ApJ...964...71H}. Right: GOODS-S field. Our galaxy candidates with  z$>$11 were found in the  JADES Medium GOODS-S region, with the grey data points again showing candidates with z$>$8  from the same study.}
    \label{fig:jades_targets}
\end{figure*}

\section{Observation and Methods}\label{data}
The debut of the JWST in 
December 2021 
opens up an unprecedented 
window into the study of the 
distant cosmos. 
Throughout the past year, 
a series of observation 
initiatives have been 
underway
e.g., JADES
\citep{2024ApJ...964...71H}, { CEERS}
\citep{2015ApJ...810...71F}, GLASS
\citep{2022ApJ...938L..15C},
NGDEEP
\citep{2023ApJ...954L..46L}
and many others. 
This paper centers on
the analysis of data 
stemming from the 
JADES programs,
with a particular focus 
on photometry.

\subsection{JADES GOODS-S and 
GOODS-N observations}
In this study, we investigate
galaxy candidates identified through 
NIRCam imaging within the GOODS-S 
and GOODS-N field.
GOODS-S
encompasses observations conducted 
under Program 1180 led by PI Eisenstein 
from UT 2022-09-29 to 2022-10-10
\citep[e.g.,][]{2023arXiv230602467B,2023arXiv230602465E,2023ApJS..269...16R,2023arXiv231012340E}. 
Additionally, NIRCam parallels, covering 
9.8 square arcminutes each, were observed 
during NIRSpec PID 1210 led by PI Ferruit 
from UT 2022-10-20 to 2022-10-24, 
positioned within and southwest of the 
JADES Medium footprint in GOODS-S. 
Further NIRCam observations (9.8 square 
arcminutes) parallel to NIRSpec PID 1286, 
also led by PI Ferruit, occurred from UT 
2023-01-12 to 2023-01-13, situated 
northwest of the JADES Deep footprint in 
GOODS-S
\citep[e.g.,][]{2023arXiv231210033R,2023MNRAS.522.6236T,2024ApJ...964...71H}. 
The total survey area for JADES GOODS-S 
spans 67 square arcminutes, with 27 
square arcminutes allocated for the JADES 
Deep program and 40 square arcminutes for 
the JADES Medium program. Utilizing 
filters F090W, F115W, F150W, F200W, 
F277W, F335M, F356W, F410M, and F444W 
($\lambda$ = 0.8 -- 5.0$\mu$m) for JADES 
Deep and the same filters without F335M 
for JADES Medium, these observations 
incorporate a minimum of six dither 
points per observation, with exposure 
times ranging from 14 to 60 ks. 
Consequently, the 5$\sigma$ depths range 
from 3.4 to 5.9 nJy, with flux aperture 
sizes varying between 1.26 and 1.52 
arcseconds. 

The NIRCam GOODS-N program encompasses a 
total area of 58 square arcminutes. 
It's split into two sections: the 
northwest (NW) covering 30.4 square 
arcminutes, and the southeast (SE) 
covering 27.6 square arcminutes. 
The NW part was observed as part of
PID 1181 (PI: Eisenstein), 
using NIRCam as the main instrument.
JADES guarantees a high level of pixel 
diversity across all filter bands, 
effectively mitigating flat-field 
inaccuracies, cosmic-ray interference, 
and other pixel-level issues. 
{ Figure 
\ref{fig:jades_targets} illustrates
the GOODS-S and GOODS-N field covered
by JADES survey, adopted from \citet{2024ApJ...964...71H}.}

For this study, we
utilized publicly available
JADES data and 
reductions\footnote{https://archive.stsci.edu/hlsp/jades}. 
We specifically used
NIRCam images from eight 
filters: F090W, F115W, F150W,
F200W, F277W, F356W, F410M, 
and F444W. 
These images were processed 
with a pixel scale of 
0.03$\arcsec$ per pixel.
To identify sources in the 
images, we created a detection
image by combining the F200W 
and F277W NIRCam bands
using a weighted sum.
We used image-segmentation tools
from 
{\tt Photutils} 
\citep{2023zndo...1035865B} 
and 
JWST data analysis
tool notebooks \footnote{https://jwst-docs.stsci.edu/jwst-post-pipeline-data-analysis}
to identify sources with a SNR
above 3 across 
five contiguous pixels. 
From this source catalog, 
we measured elliptical Kron 
aperture fluxes using
{\tt Photutils}.

We used a flexible Kron
scale factor (K = 1.5) and
a circularized radius six
times larger than the 
Gaussian-equivalent elliptical 
sizes to ensure accurate
measurements while masking 
any segmentation regions of 
neighboring sources
\citep{2023ApJ...955...13B,2024ApJ...964...71H}.
The size of elliptical
Kron
aperture for each source was
estimated by
multiplying the Kron scale factor by the
individual Kron radius. 

We 
extracted and fitted to SED for 
each source adopting the Kron
fluxes,  as previous studies
have demonstrated that
measuring colors in small
elliptical apertures 
accurately captures the 
colors of distant galaxies
\citep{2023ApJ...946L..13F,2023ApJ...955...13B}. 
Additionally, we repeated 
a similar analysis 
using Kron apertures with
a scale factor of 2.5. 
To correct for aperture effects, 
we estimated an aperture
correction by comparing the 
total fluxes measured in the larger
and smaller apertures for
each source.
This correction factor was 
then applied to the fluxes 
and uncertainties for all
filters.

\subsection{Photometric redshift estimation}\label{sec:eazypy}
We used \texttt{EAZY}-py, the Python 
implementation of the \texttt{EAZY}
photometric redshift estimator code 
\citep{2008ApJ...686.1503B},
to derive the initial photometric
redshifts.
We utilize a set of 12 
``tweak$\_$fsps$\_$QSF$\_$12$\_$v3''
templates derived from the 
Flexible Stellar Population Synthesis 
(FSPS) library 
\citep{2009ApJ...699..486C,2010ascl.soft10043C}.
These templates include a range of galaxy 
types including star-forming, quiescent, 
and dusty and also include a range of 
star-formation histories, e.g., bursty, 
slowly rising, and slowly falling.
Additionally, we utilize
six newly developed templates 
from \citet{2023ApJ...958..141L}, 
which were also applied by 
\cite{2024ApJ...965..169A} for 
deriving photometric redshifts of
galaxies with z > 7.5.
These additional templates were
found
to provide better photometric redshift 
estimation for $z$ $>$ 9 galaxies.

We utilized the $\chi^2(z)$ values 
generated by \texttt{EAZY} to compute a
probability distribution, P($z$),
assuming a uniform redshift prior. 
This distribution is given by the 
equation $P(z) = {\rm exp}[\chi^2(z)/2]$, 
and it is normalized such that the 
integral of P($z$) over all redshifts 
equals 1.0. 
We vary redshifts in 0.1 $<$ $z$ $<$ 20, in 
steps of 0.01. 
Since the high-redshift luminosity funciton
is still not well-constrained, we adpoted
a flat luminosity
prior to prevent bias against the selection
of bright high-redshift
galaxies. Similar techniques were 
also
adopted in several past JWST studies,
such as \citet{2022ApJ...940L..55F,2023ApJ...955...13B}.

\section{Selection criteria for high-redshift candidates}\label{selection}

We selected high-redshift galaxy candidates
by combining \texttt{EAZY}-based SED fitting 
for photometric redshifts,
signal-to-noise
(SNR) criteria, as adopted in several
past studies such as 
\citet{2020ApJ...889..189S,2022ApJ...940L..55F,2023ApJ...946L..13F,2023ApJ...955...13B, 2023ApJ...954L..46L, 2023MNRAS.519.1201A,2023MNRAS.518.6011D,2024ApJ...964...71H}.
We adopted photometric SNR criteria to
ensure nondetections in filters bluer
than Ly$\alpha$ and robust detections in 
filters redder than Ly$\alpha$ break.
Finally, we visually inspect each 
galaxy candidates for detector artifacts, 
diffraction spikes, and spurious noise close
to detector edge.

Our selection criteria for identifying 
galaxy candidates with $z$ $>$ 11 are 
similar to the criteria adopted in
\citet{2022ApJ...940L..55F} and 
\citet{2023ApJ...955...13B}-

\begin{enumerate}

    \item The best-fit photometric redshift,   derived from \texttt{EAZY} SED fitting, must exceed 11 (i.e., $z$ $>$ 11).

    \item An SNR of $<$ 1.5 to ensure non-detections blueward of Lyman-$\alpha$ in all of the following filters F090W and
    F115W for $z$ $<$ 13. An additional
    F150W SNR of $<$ 1.5 was also adopted for 
    $z$ $\geq$ 13.

    \item An SNR $>$ 5.5 in at least two of the following filters: F150W, F200W, F277W, F356W, F410M, or F444W for $z$ $<$ 13 and F200W, F277W, F356W, F410M, or F444W for $z$ $>$ 13.

    \item  The integral of the \texttt{EAZY}
    posterior redshift probability (P($z$)) 
of the galaxy being at z$>$11 is greater than 70\%, i.e. $\int_{z=11}^{\infty} P(z) \, dz > 0.7$.

    \item  The angular separation 
between the candidates and 
any other 
nearby bright sources
are at least 0.3$\arcsec$, 
corresponding to 10 pixels.
\end{enumerate}

In addition to the above criteria, we
fitted each SED twice, first restricting
\texttt{EAZY} maximum redshift to $z$ $<$ 7 
(low-redshift)
and then varying the redshift between 
0 $<$ $z$ $<$ 20 (high redshift; see Figure \ref{fig:snapshot0}--\ref{fig:snapshot06}). 
We estimated the disparity between
the best-fit $\chi^2$ for the low-redshift 
and high-redshift fit. For this work,
we only consider $\Delta\chi^2$ $\geq$ 9,
ruling out the low-redshift model
at $\geq$ 3$\sigma$ significance
\citep{2023ApJS..265....5H}.

\section{Results and Discussion}\label{results}
Adopting the selection criteria and
SED fitting method, we detect six new galaxy candidates at
redshift $z$ $>$ 11 in the JADES GOODS-S and
GOODS-N field. 
Below we discuss
our primary results.

\subsection{Galaxy sample with $z$ $>$ 11}
Utilizing fits generated by \texttt{EAZY}, we identify six 
new candidates within the redshift range
of 11 $<$ $z$ $<$ 15  across the GOODS-S
and GOODS-N 
fields.
Specifically, the five candidates
are detected in GOODS-N, with the 
remaining two candidates detected 
in the GOODS-S field.
Among the five candidates in GOODS-N,
three are located in the southeastern 
sector, while the remaining two are 
located in its northwestern region.
The two targets found in the GOODS-S field are specifically located within its medium region.  Figure \ref{fig:jades_targets} 
illustrates the positions 
of these candidates in the GOODS-N 
and GOODS-S fields.  
Among the six candidates, three had 
redshifts between 11 and 12, while the 
remaining four showed best-fit photometric 
redshifts exceeding 12. 
Table \ref{t:bagpipes} lists the newly 
identified candidates along with best-fit
parameters from SED fitting.

\begin{figure*}
    \centering
        \begin{tabular}{c}
\includegraphics[width=0.9\textwidth]{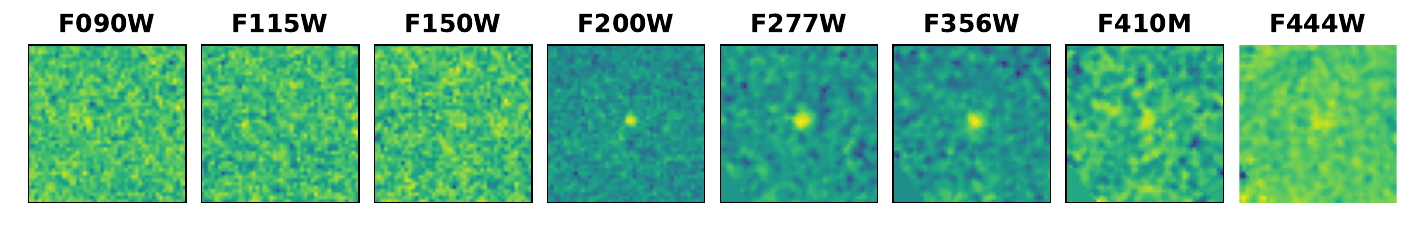}  \\
    \end{tabular}
    \hspace{-20pt}
    \begin{tabular}{cc}
    \includegraphics[width=0.6\textwidth]{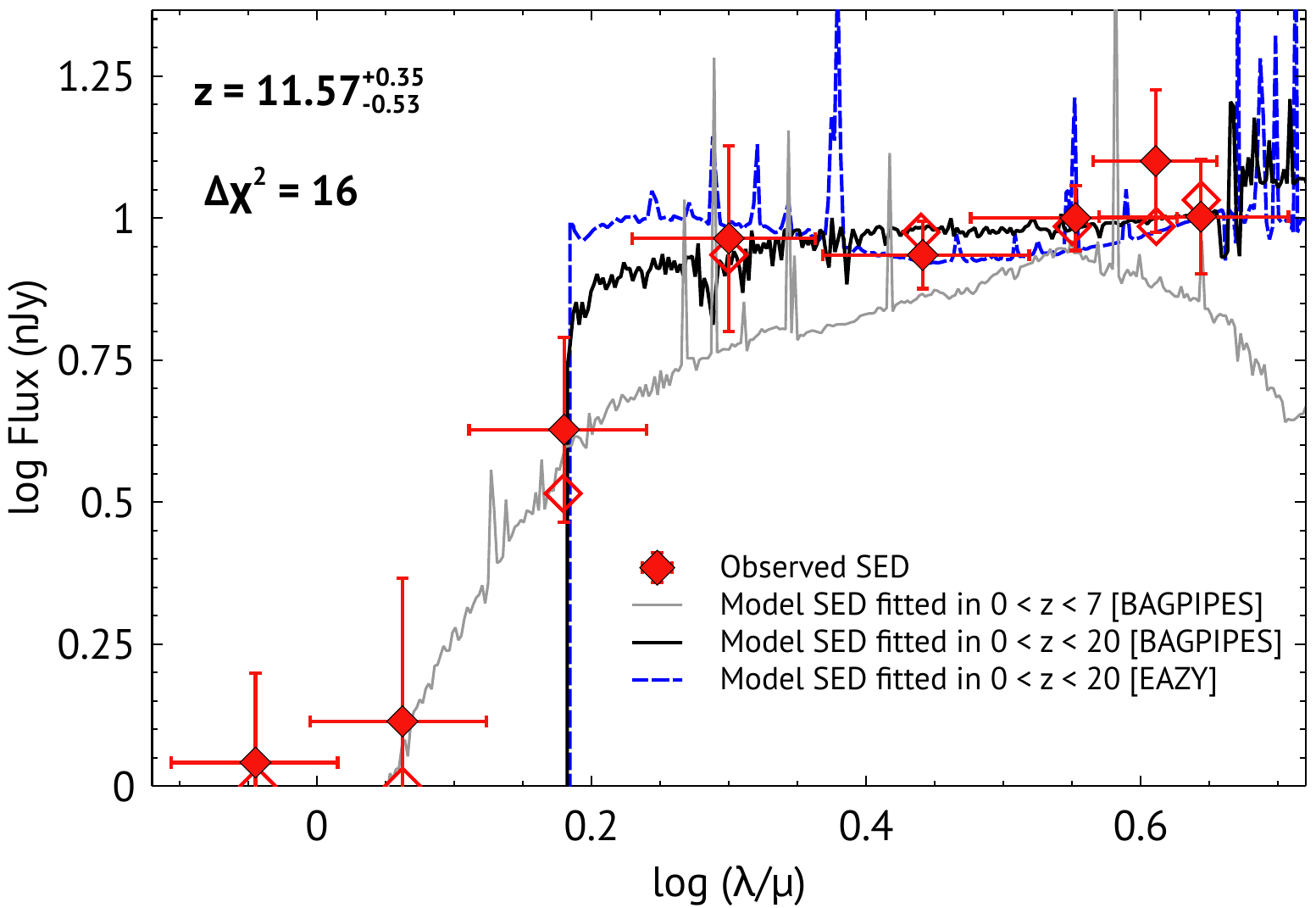}  &
\includegraphics[width=0.3\textwidth]{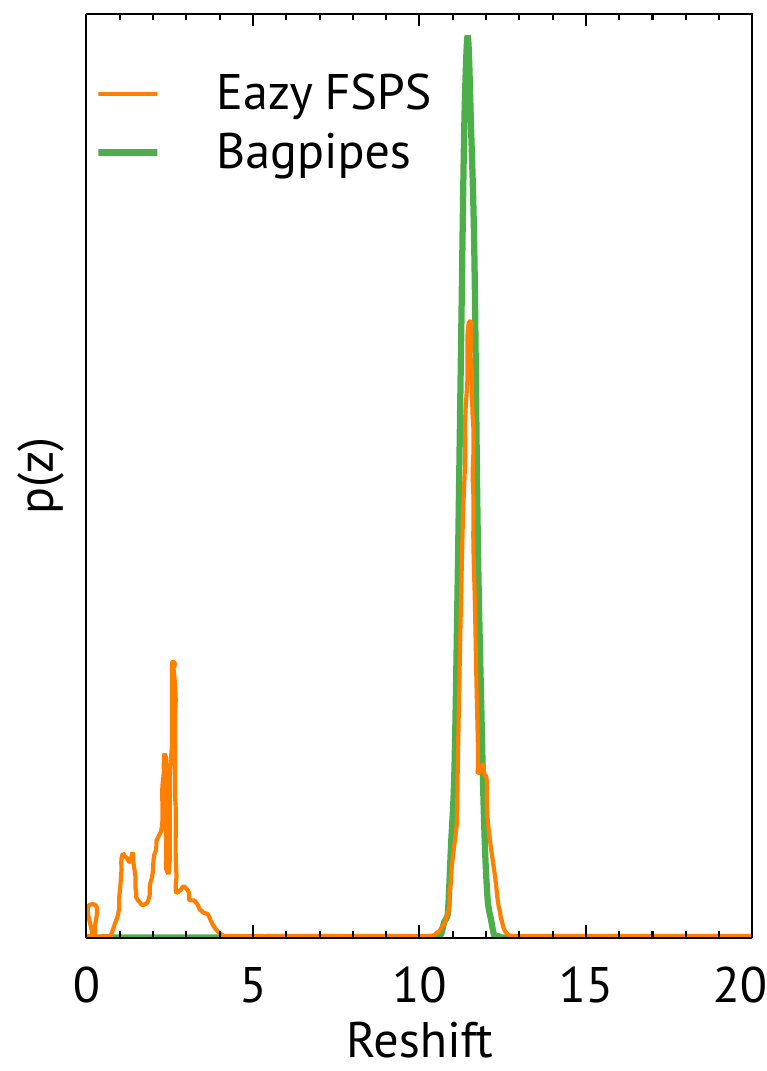}  \\
    \end{tabular}
    \caption{ Thumbnail images for JADES-- 53.16238 -27.87016. In each panel, the top rows display 1.8$\arcsec$ $\times$ 1.8$\arcsec$ image cutouts for the 8 JWST filters. The bottom-left panels represent the observed photometry in red data points along with their corresponding 1$\sigma$ uncertainties. 
    The best-fit SEDs from \texttt{BAGPIPES} for the preferred high-redshift solution are shown as a black
    solid curve, while the 
    corresponding
    SEDs for the low-redshift solution are represented as a 
    solid grey curve. 
    { Open red diamonds represent best-fit \texttt{BAGPIPES} model
    photometry for high-redshift solution.}
    The best-fit SEDs with 
    {\tt EAZY} for the preferred
    high-redshift solution are shown 
    with blue dashed curve.
    The redshift range displayed in the plot was determined using \texttt{EAZY}. In addition, 
    the posterior probability distributions, P($z$), for photometric redshift solutions is displayed with the orange curves in the bottom-right panels.}
    \label{fig:snapshot0}
\end{figure*}
\begin{figure*}
    \centering
        \begin{tabular}{c}
\includegraphics[width=0.9\textwidth]{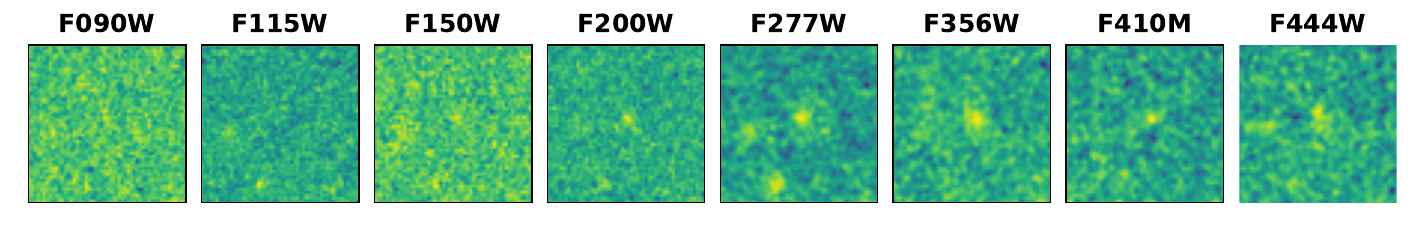}  \\
    \end{tabular}
    \hspace{-20pt}
    \begin{tabular}{cc}
    \includegraphics[width=0.6\textwidth]{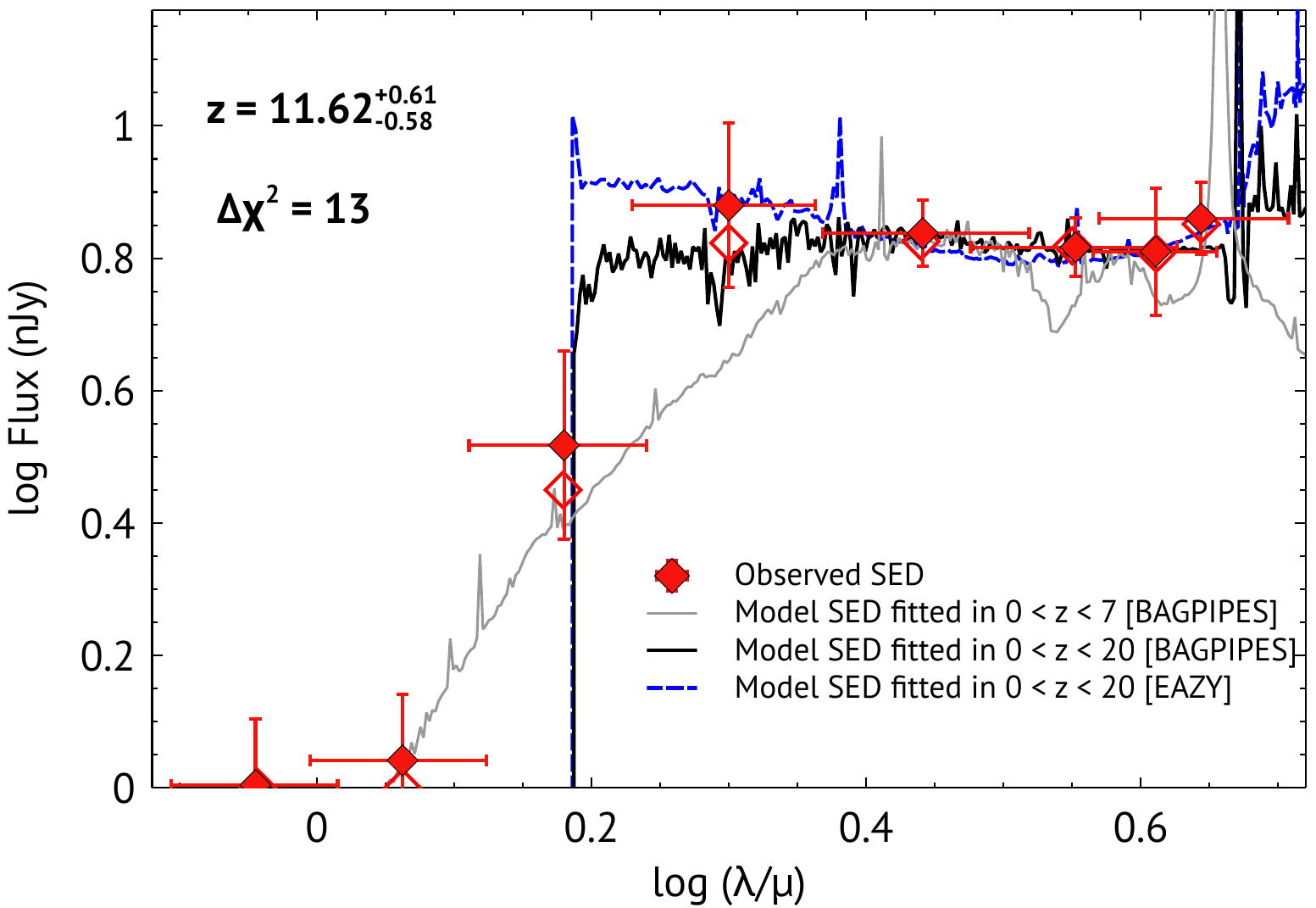}  &
\includegraphics[width=0.3\textwidth]{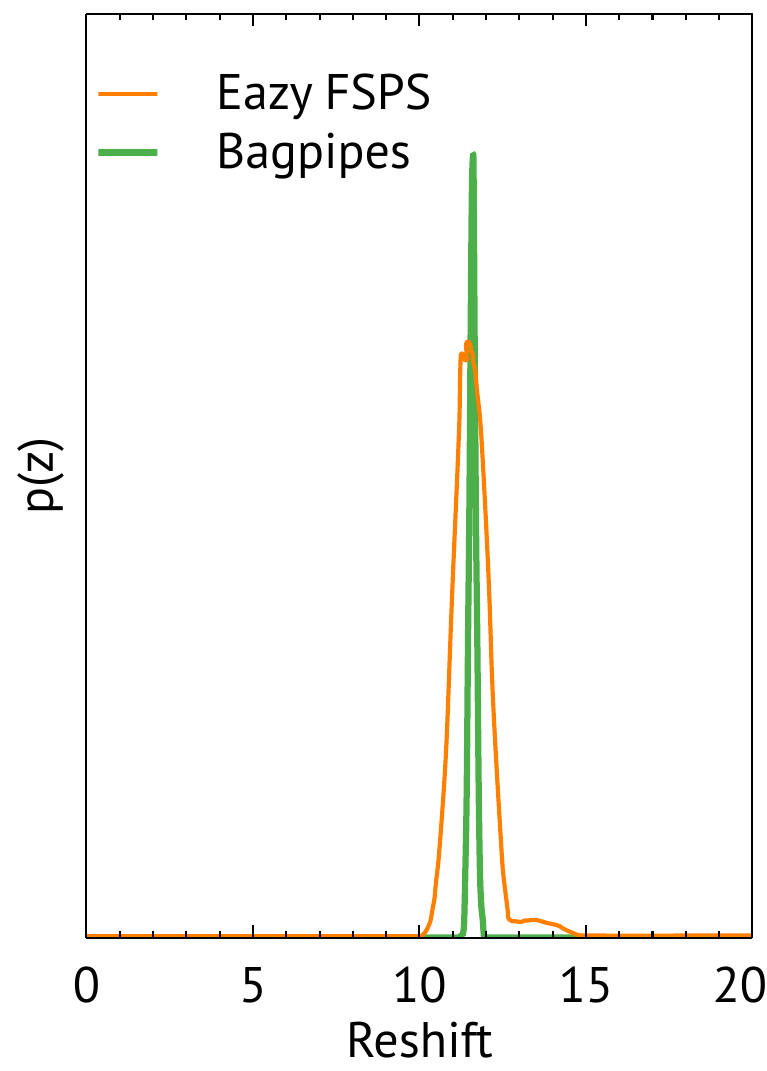}  \\
    \end{tabular}
    \caption{Same as Figure \ref{fig:snapshot0}, but for JADES-- 189.17805+62.18788.}
    \label{fig:snapshot01}
\end{figure*}
\begin{figure*}
    \centering
        \begin{tabular}{c}
\includegraphics[width=0.9\textwidth]{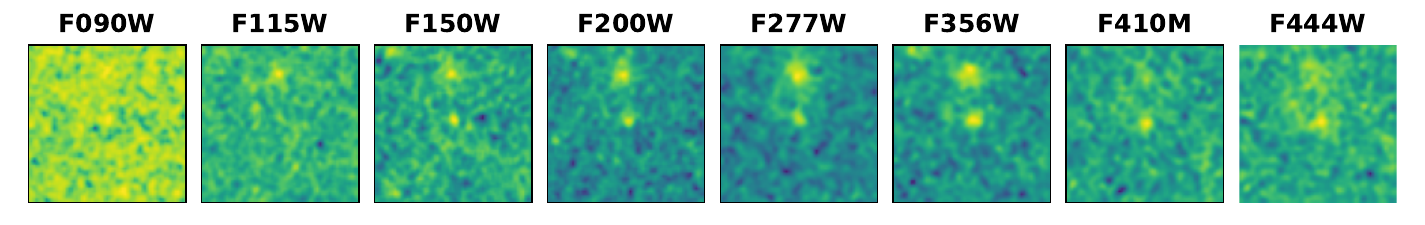}  \\
    \end{tabular}
    \hspace{-20pt}
    \begin{tabular}{cc}
    \includegraphics[width=0.6\textwidth]{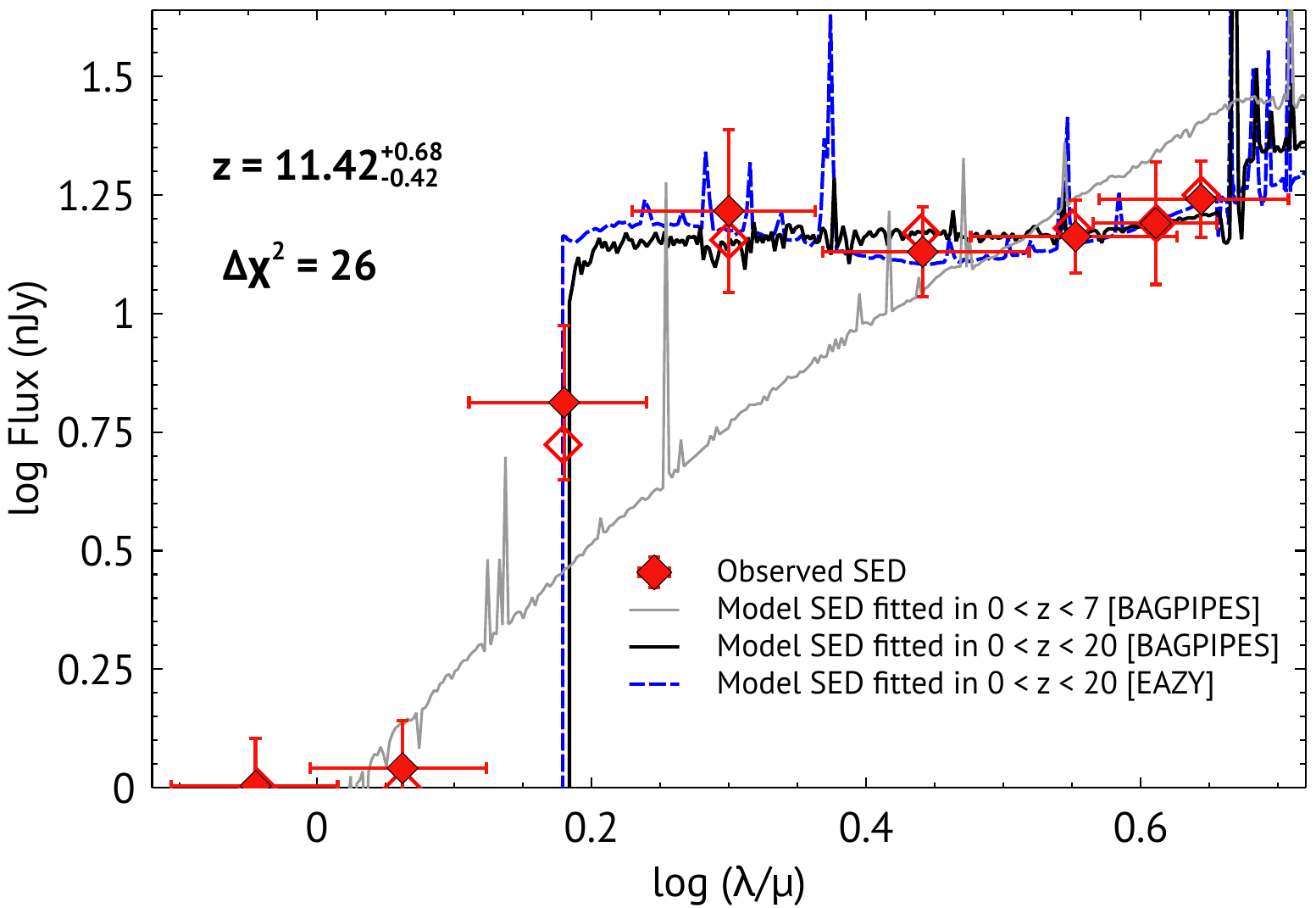}  &
\includegraphics[width=0.3\textwidth]{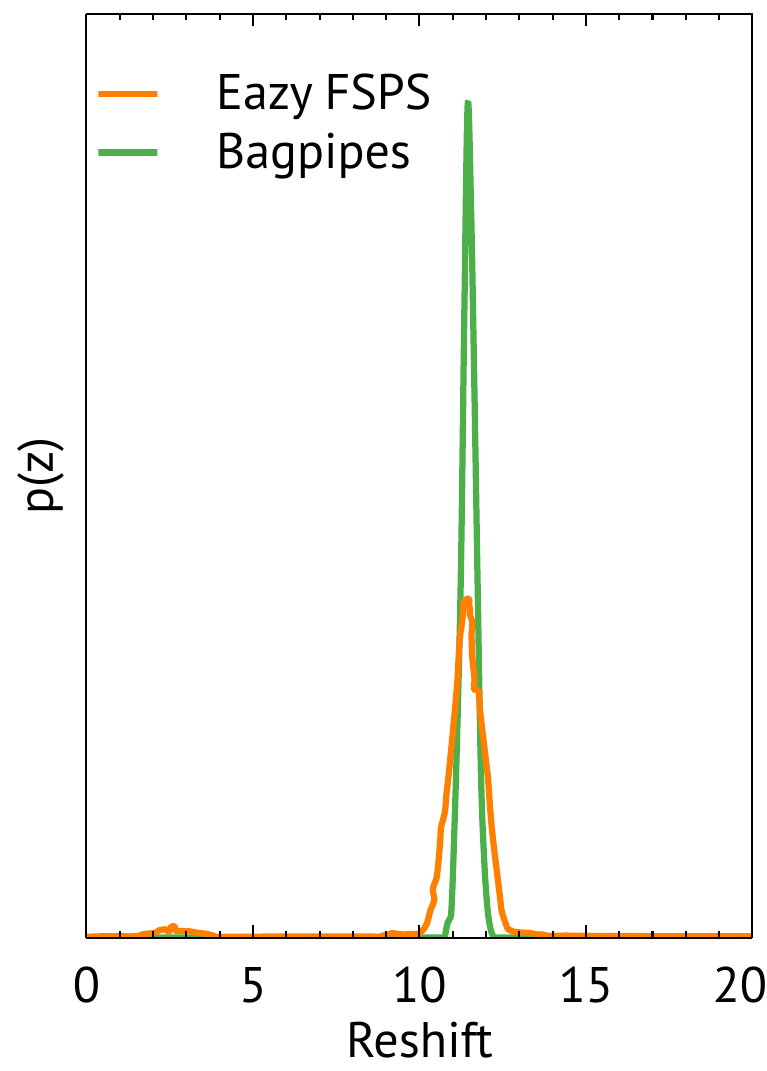}  \\
    \end{tabular}
    \caption{Same as Figure \ref{fig:snapshot0}, JADES-- 189.09621+62.24877}
    \label{fig:snapshot02}
\end{figure*}
\begin{figure*}
    \centering
        \begin{tabular}{c}
\includegraphics[width=0.9\textwidth]{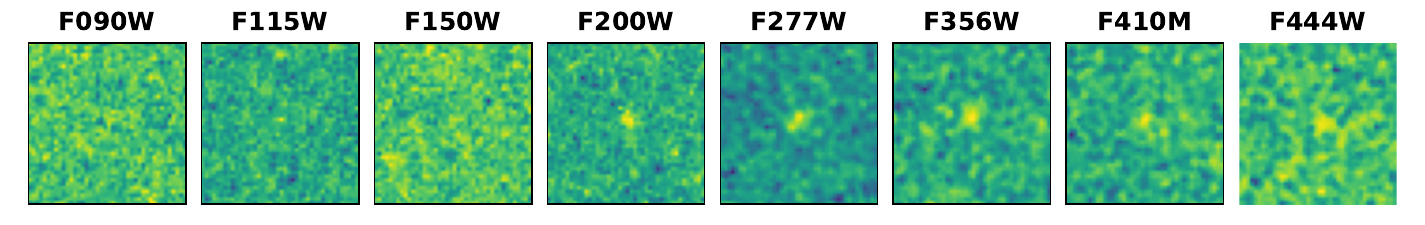}  \\
    \end{tabular}
    \hspace{-20pt}
    \begin{tabular}{cc}
    \includegraphics[width=0.6\textwidth]{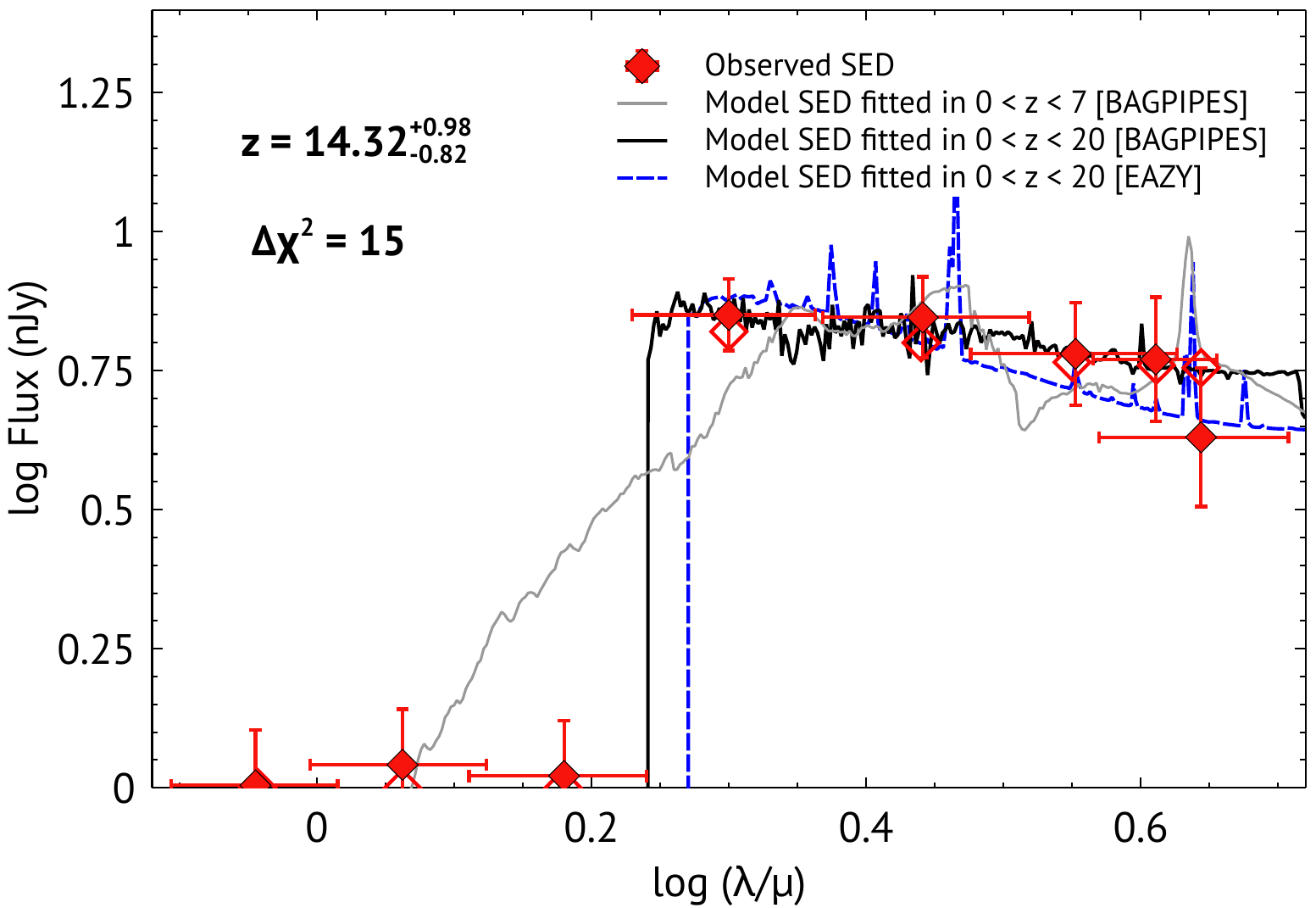}  &
\includegraphics[width=0.3\textwidth]{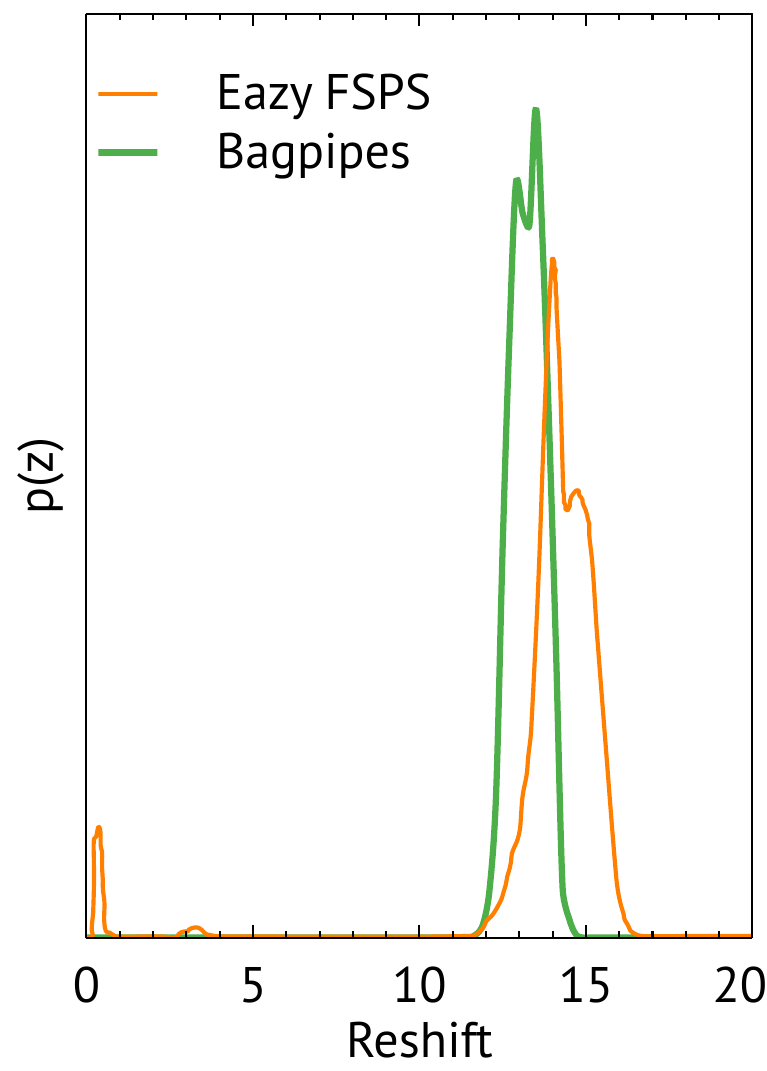}  \\
    \end{tabular}
    \caption{Same as Figure \ref{fig:snapshot0}, but for JADES-- 189.29834 +62.21367}
    \label{fig:snapshot04}
\end{figure*}
\begin{figure*}
    \centering
        \begin{tabular}{c}
\includegraphics[width=0.9\textwidth]{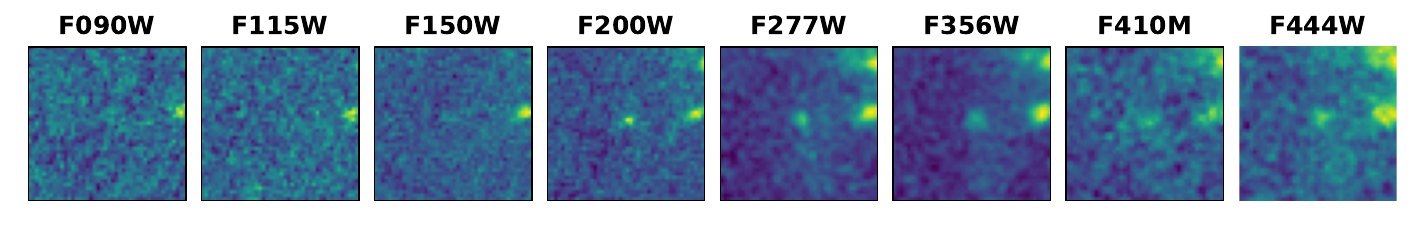}  \\
    \end{tabular}
    \hspace{-20pt}
    \begin{tabular}{cc}
    \includegraphics[width=0.6\textwidth]{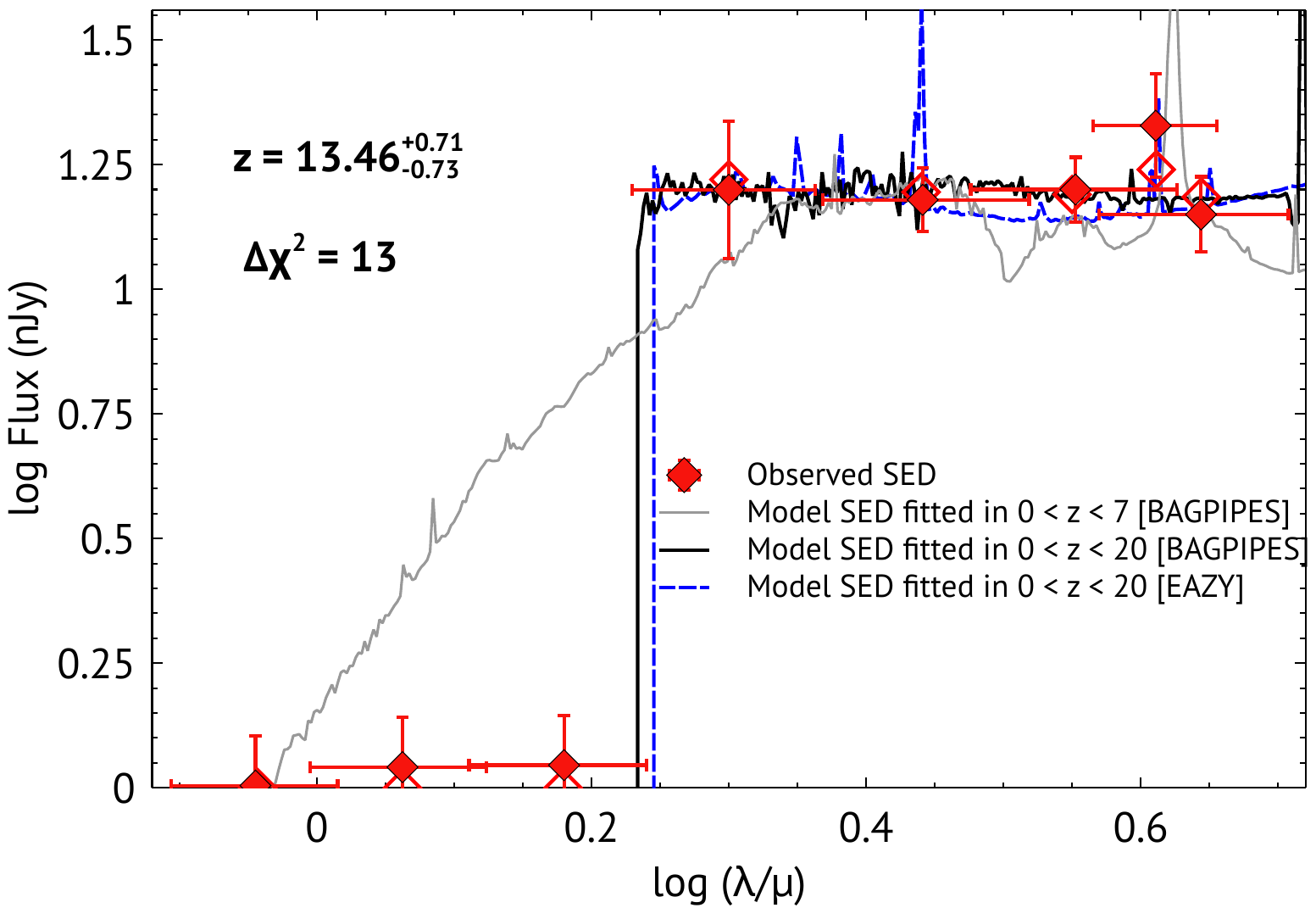}  &
\includegraphics[width=0.3\textwidth]{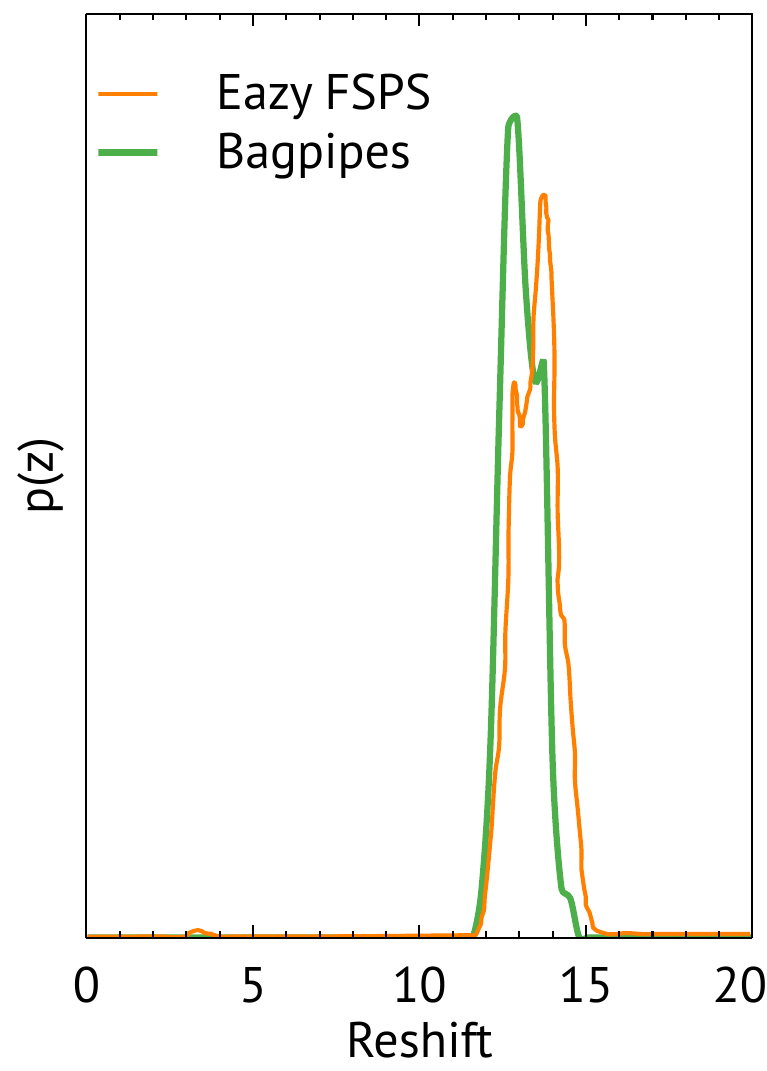}  \\
    \end{tabular}
    \caption{Same as Figure \ref{fig:snapshot0}, but for JADES-- 189.10753+62.24305.}
    \label{fig:snapshot05}
\end{figure*}
\begin{figure*}
    \centering
        \begin{tabular}{c}
\includegraphics[width=0.9\textwidth]{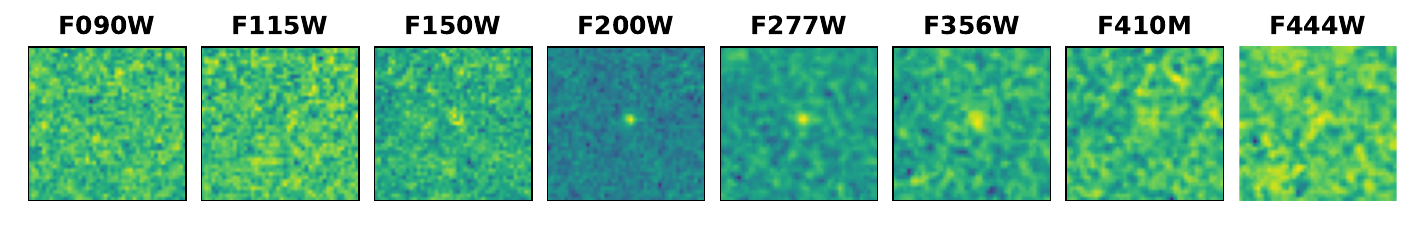}  \\
    \end{tabular}
    \hspace{-20pt}
    \begin{tabular}{cc}
    \includegraphics[width=0.6\textwidth]{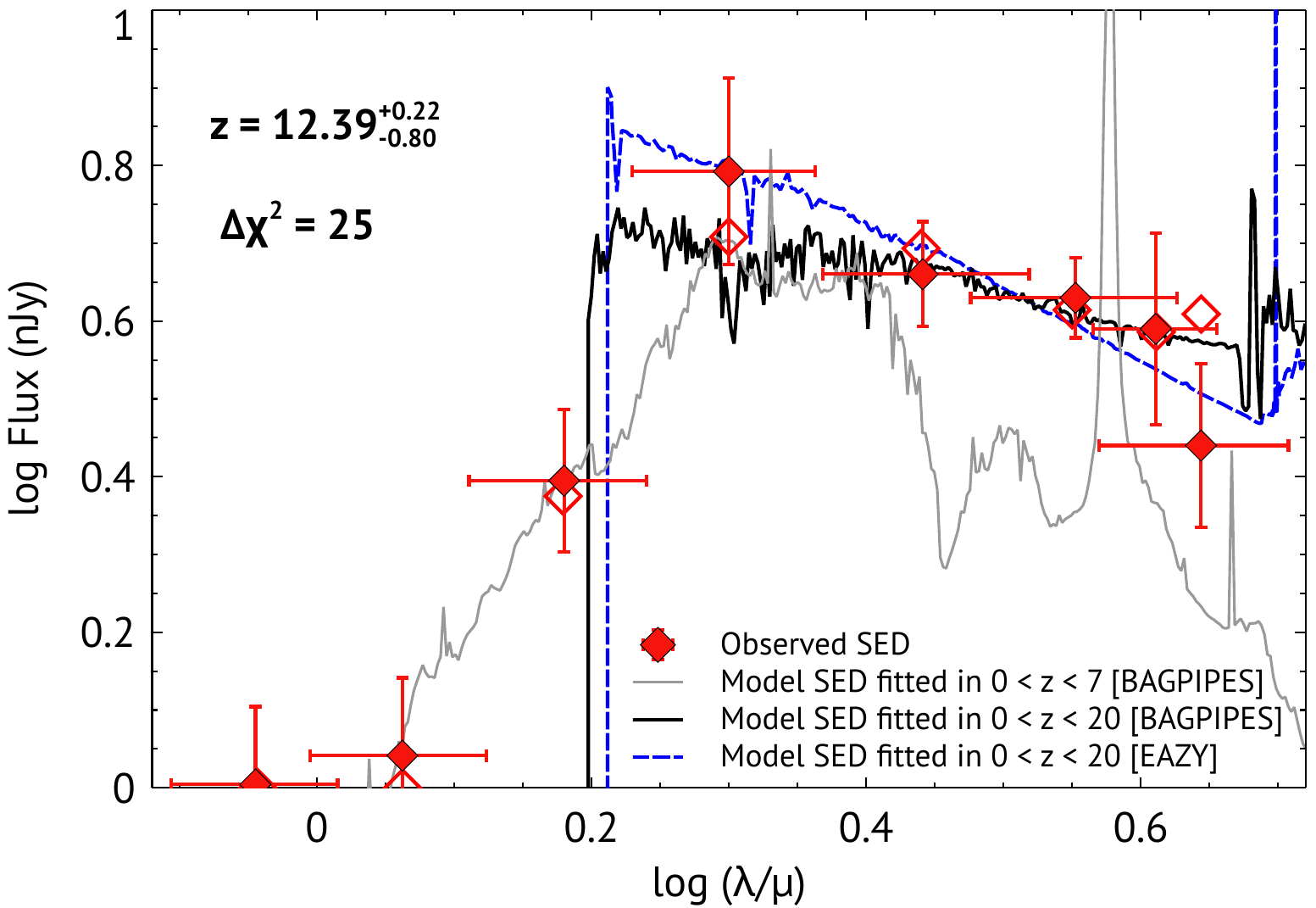}  &
\includegraphics[width=0.3\textwidth]{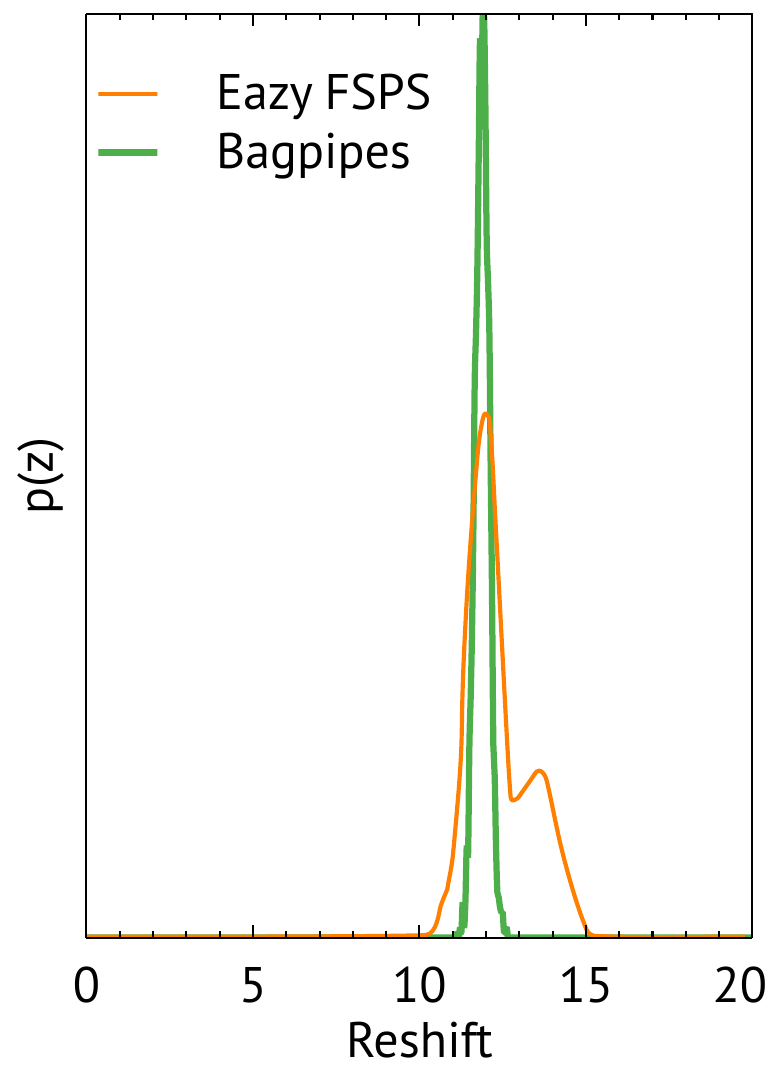}  \\
    \end{tabular}
    \caption{Same as Figure \ref{fig:snapshot0}, but for JADES-- 189.26307+62.20648.}
    \label{fig:snapshot06}
\end{figure*}

All candidates are robust detections 
characterized by a precise high-redshift 
solution without any significant competing low-redshift 
alternatives, with $\sim$ 70-100\% of their 
total probability concentrated within 
$\Delta$z = 1 around the best-fit solution. 
Similar quality classification criteria have 
been adopted in a recent \textit{JWST}
UNCOVER 
high-redshift survey 
\citep{2023MNRAS.524.5486A}, 
where candidates meeting such criteria
have been labeled as high-quality.
Our sample targets the faint end of the 
luminosity distribution,  
covering a magnitude range of
m$_{F277W}$ =  28.5 and 30.4.
Previous JADES surveys have documented 
systems with F277W AB magnitudes reaching
as faint as 31.0 \citep{2024ApJ...964...71H}.
Figure \ref{fig:snapshot0}, \ref{fig:snapshot01}, \ref{fig:snapshot02},  \ref{fig:snapshot04}, \ref{fig:snapshot05}, and \ref{fig:snapshot06}
 shows the {1.8$\arcsec$ $\times$ 
1.8$\arcsec$} thumbnail images, 
best-fit SEDs from \texttt{BAGPIPES} 
and {\tt EAZY}, 
and the posterior redshift distributions.
The red points represent the measured 
photometry. 
The best-fit favored high-redshift
solutions (represented by black
solid curves) 
have been shown along with the best-fit
low-redshift solutions (represented by 
grey solid curves).  
The SED modeling and the physical properties of the galaxies
derived from \texttt{BAGPIPES} fits 
are described in Section \ref{properties}.

In our dataset, we identified 
three targets within the lowest 
redshift range in our sample, 
11 $<$ z $<$ 12  (\texttt{EAZY} determined), as illustrated
in Figure \ref{fig:snapshot0},  \ref{fig:snapshot01}, and \ref{fig:snapshot02}: 
JADES--53.16238-27.87016 (GOODS-S),
JADES-- 189.17805+62.18788 (GOODS-N), 
and JADES-- 189.09621+62.24877 (GOODS-N). 
In the observed $11<z<12$ galaxies, 
none exhibit substantial flux in the F115W images, 
thus classifying them as F115W dropout candidates.
In the next higher redshift range, 
our 
survey identified 
three more candidate galaxies 
with photometric redshifts $12<z<15$  (\texttt{EAZY} determined):
JADES--189.26307+62.20648 (GOODS-N), 
JADES-- 189.29834 +62.21367    (GOODS-N), and JADES-- 189.10753+62.24305   (GOODS-N).
The corresponding thumbnail images are 
displayed in Figure  \ref{fig:snapshot04}, \ref{fig:snapshot05},
and \ref{fig:snapshot06}. 
JADES-- 189.26307+62.20648 is categorized as an F115W dropout. The $z\geq13$ targets, JADES-- 189.29834+62.21367  and JADES-- 189.10753+62.24305  are both categorized as F150W dropouts.

\subsection{Size of galaxies}

\begin{figure}
    \centering
    \hspace{-20pt}
\includegraphics[width=0.5\textwidth]{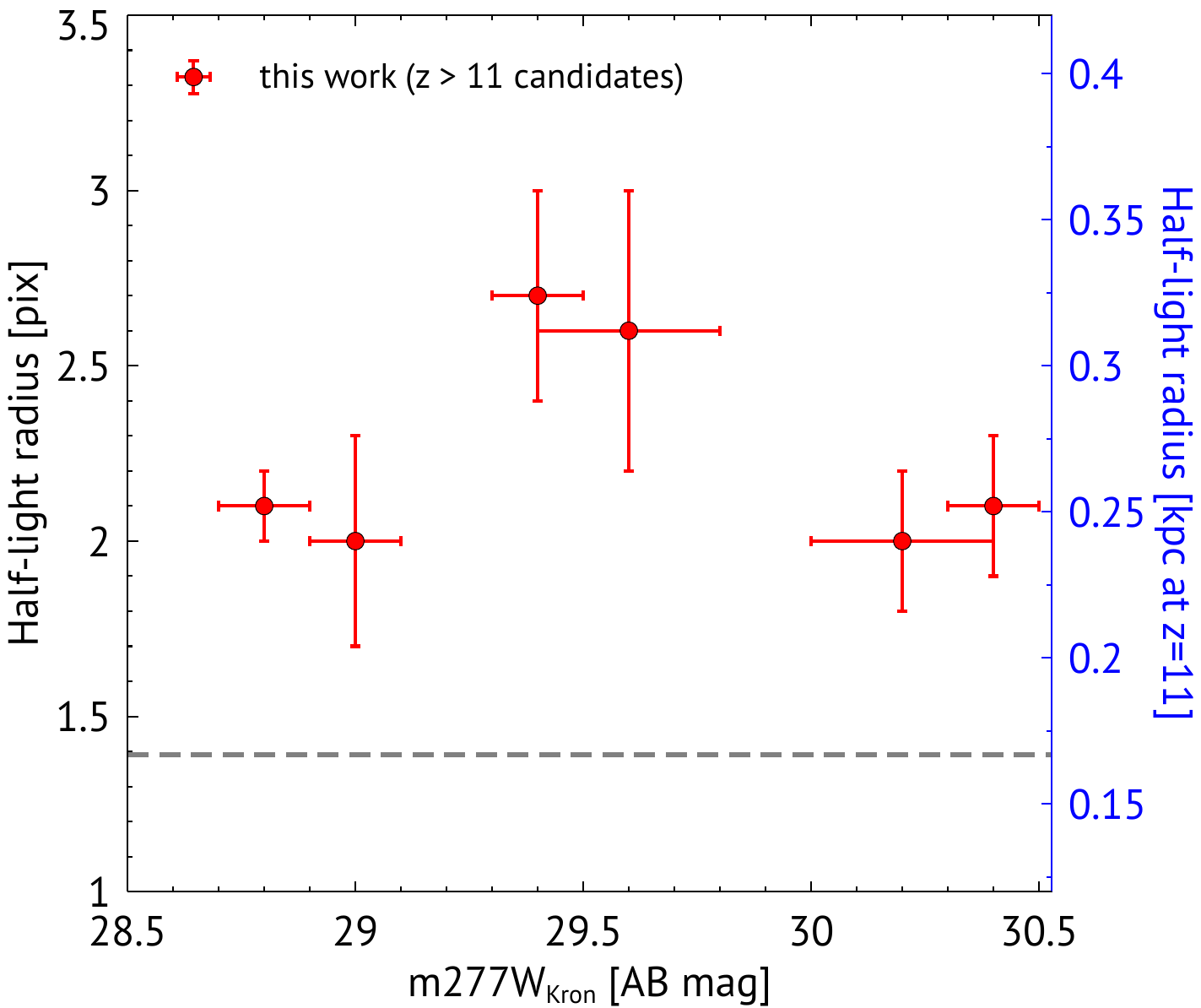}  \\
\caption{Half-light radius vs. AB Kron magnitude
in F277W filter. Red data points show
the Half-light radius of our sample galaxy
candidates with 1$\sigma$ uncertainties.
A secondary y-axis for
half-light radius in kpc (scaled
at $z$ = 11) is shown on the right side. The grey dashed line displays the 
F277W PSF FWHM in pixels (1.397 pixels).}
\label{fig:f-size}
\end{figure}

We estimated the sizes of all 7 galaxies in our sample through the parametric fitting of the NIRCam images using the Python package \texttt{Photutils}, an Astropy package for photometry and detection of astronomical sources \citep{2022zndo...6825092B}.
We estimated the half-light radius for F277W filter using the Kron scale factor of 2.5, as adopted by \citet{2023ApJ...955...13B}.\\
In Figure \ref{fig:f-size}, we present the F277W half-light radii of our high-redshift galaxy candidates.
The figure displays both the pixel size and the corresponding physical sizes (in kpc). The physical sizes were estimated using the photometric redshifts derived from \texttt{BAGPIPES} fits, as shown in Table \ref{t:bagpipes}. We find half-light radii ranging between 2.0 pixels to 2.7 pixels, corresponding to an angular spread between
0.06$\arcsec$ and 0.081$\arcsec$ (adopting pixel size of 0.03$\arcsec$), with a median of 0.06 $\arcsec$. Given the F277W PSF full-width at half maximum of 0.088$\arcsec$, and a radius of 0.044$\arcsec$, all the galaxies are spatially resolved. The angular spreads translate to half-light radii ranging between 0.24 kpc to 0.34 kpc, with a median of 0.25 kpc for our galaxy candidates.
 These dimensions are in agreement with the rest-UV sizes reported in the GLASS survey, indicating a median half-light radius of 
 0.45 $\pm$ 0.13 kpc for 7$<$z$<$15 galaxies in the F444W band \citep{2022ApJ...938L..17Y}, as well as with the CEERS survey, which found a median half-light radius of 0.46 kpc for 9$<$z$<$16 galaxies in the F200W band \citep{2023ApJ...946L..13F}.

\subsection{Physical Properties through SED fitting}\label{properties}

To derive the physical properties of the 
identified galaxy candidates, we utilized the 
state-of-the-art Bayesian SED fitting code 
\texttt{BAGPIPES} \citep{2018MNRAS.480.4379C} 
to fit the observed SED. \texttt{BAGPIPES} 
generates complex model galaxy spectra and fits 
photometric and spectroscopic observations 
using 
{ nested sampling algorithm 
\citep[e.g., nautilus sampler;][]{2024OJAp....7E..79D}}, thereby generating 
posterior distributions of galaxy properties 
for each source in the sample.
\texttt{BAGPIPES} is capable of modeling 
galaxies with various star formation histories 
(SFHs), including delayed-$\tau$, constant, and 
bursts \citep{2020ApJ...904...33L}. The stellar 
population synthesis models adopted for this 
work are based on BC03 
\citep{2003MNRAS.344.1000B} version 2016. These 
models assume the \citet{2002Sci...295...82K} 
initial mass function (IMF) and include nebular 
line and continuum emission using $\Cloudy$ 
\citep{2023RMxAA..59..327C}.
For this study, we selected a delayed-$\tau$ 
SFH (where SFR($t$) $\propto$ $t e^{-t/\tau}$)
{ following \citet{2024ApJ...964L..24M}}, known for its effectiveness in 
modeling the majority of the stellar population 
and accurately estimating stellar masses 
\citep{2017A&A...608A..41C}.

We modified the \texttt{BAGPIPES} 
source code to expand the default
redshift range from $z$ = $0-10$ to 
$z$ = $0-20$,  
ensuring consistency with the \texttt{EAZY}-py 
results used to select the galaxy sample
(Section \ref{sec:eazypy}). 
The \texttt{BAGPIPES} 
SED-fitting was conducted over a 
wide range of parameter space. 
We vary $\tau$ within 0.1 and 14 Gyr, 
stellar masses within $\log (M_*/M_\odot)$= 
4 and 13, and log of metallicities
(log Z/Z$_{\odot}$) within
0.005 and 2.5. The logarithm of the
ionization parameter (log U),
for accounting nebular line and continuum
emission, 
was allowed to vary within $-4$ and $-2$.
For this work we adopted 
Calzetti dust attenuation
curve to reduce the number
of free parameters in our fits
\citep{2000ApJ...533..682C}. 
We let the dust extinction $A_V$ vary
within 0
and 4. 
Additionally, we introduced a 
multiplicative factor (1 $<$ $\eta$ $<$ 2) 
to the dust model to address birth-cloud
dust attenuation,
which typically doubles around $\hii$
regions compared to the general ISM
within the galaxy's first 10 Myr 
\citep{2023ApJ...955...13B}.
To account for this, we capped the
maximum age of the birth-cloud at 0.01 Gyr
\citep{2023ApJ...948..126G}.
The age parameter was allowed to range
from 1 Myr to the age of the Universe.
Following a similar approach to the 
\texttt{EAZY} fitting, we conducted two fits 
of the observed photometry with 
\texttt{BAGPIPES}.
Initially, we constrained the redshift
range to $z$ $<$ 7. 
Then, we performed another fitting 
while allowing the redshift to
vary between 0.1 and 20.
Figure \ref{fig:snapshot0},  \ref{fig:snapshot01}, \ref{fig:snapshot02},
\ref{fig:snapshot04}, \ref{fig:snapshot05}, and \ref{fig:snapshot06} 
display the best-fit model SEDs obtained 
from the \texttt{BAGPIPES} fitting, 
accompanied by the corresponding
$\Delta\chi^2$ values.

The physical parameters extracted
from the \texttt{BAGPIPES} fittings
are detailed in Table \ref{t:bagpipes}.
Among these parameters, the inferred
intrinsic stellar masses exhibit a 
range approximately from 
$\log M_\ast/M_\odot \sim 7.75$ to 8.75,
while the corresponding star formation
rates (SFRs) span from around 
$\sim 0.9$ to 5.1 $M_\odot/\text{yr}$. 
Additionally, the specific star
formation rates (sSFRs), quantified 
as $\log \text{sSFR}/\text{Gyr}$, 
vary within the range of approximately 
$\sim 0.95$ to 1.46. 
Notably, similar sSFRs have been
found for galaxy candidates at 
redshifts $z > 8$ in \citet{2023ApJ...955...13B}.

\begin{table*}
    \centering
 \caption{ Photometric redshifts from \texttt{EAZY} and \texttt{BAGPIPES}
 SED fitting, along with physical properties of the galaxy candidates estimated with \texttt{BAGPIPES}. 
 Best-fit parameters are correspond to
 delayed-$\tau$ star formation history
 model, as presented in Figures 
 \ref{fig:snapshot0}--\ref{fig:snapshot06}.
 Errors are estimated with 1$\sigma$ uncertainties.\\
 $^{\ast}$ -- Redshifts as listed in JADES catalog.\\
 $^{\dagger}$ -- Best-fit redshifts
 from {\tt EAZY} fit within $0<z<20$.\\
  $^{\dagger\dagger}$ -- Best-fit redshifts
 from {\tt BAGPIPES} fit within $0<z<20$.\\
 \label{t:bagpipes}}   
 \setlength{\tabcolsep}{5pt}
\scalebox{0.85}{
\renewcommand{\arraystretch}{1.}
\centering                          
\begin{tabular}{lccccccccccccc}
\hline
\hline
\ \ RA & DEC & z$_{\rm JADES}^{\ast}$ & z$_{\tt EAZY}^{\dagger}$ & z$_{\tt Bagpipes}^{\dagger\dagger}$ & m$_{\rm F277}$  & $\Delta \chi^2$ &  $P(z)$ & $\log (M_*/M_\odot)$ & SFR (100 Myr) & SFR (10 Myr) & Age & Av & $r_{hl}$\\ 
\ \ (deg) & (deg)& & ($z\_a$) & & (AB) & (68\%)& & & $(M_{\odot}\text{yr}^{-1})$ & $(M_{\odot}\text{yr}^{-1})$ & (Myr) & & (pix/arcsec)\\
\hline
GOODS-S & & & & & & & & & & \\\\

\ \ {53.16238} & -27.87016 & 3.5 & 11.57$^{+0.35}_{-0.53}$ & 11.46$^{+0.24}_{-0.23}$ & 28.8 $\pm$ 0.1 & 16 & 0.7 & 8.59$^{+0.19}_{-0.29}$ & 4.2$^{+2.2}_{-1.1}$ & 2.4$^{+1.4}_{-0.5}$ & 200$^{+130}_{-120}$ & 0.45$^{+0.15}_{-0.15}$ & 2.1 $\pm$ 0.1\\\\


\hline
GOODS-N & & & & & & & & \\\\

\ \ 189.17805 & 62.18788 & 2.91 & 11.62$^{+0.61}_{-0.58}$ & 11.5$^{+0.1}_{-0.1}$ & 30.4 $\pm$ 0.1 & 13 & 0.9 & 8.25$^{+0.17}_{-0.27}$ & 1.8$^{+0.7}_{-0.3}$ & 1.7$^{+0.7}_{-0.5}$ & 230$^{+100}_{-140}$ & 0.22$^{+0.12}_{-0.1 1}$ & {2.1 $\pm$ 0.2} \\\\

\ \ 189.26307 & 62.20648 & 9.51 & 12.39$^{+0.22}_{-0.80}$ & 11.89$^{+0.20}_{-0.22}$ & 29.0 $\pm$ 0.1 & 25 & 0.9 & 7.75$^{+0.26}_{-0.31}$ & 0.9$^{+0.2}_{-0.2}$ & 0.85$^{+0.20}_{-0.13}$ & 120$^{+80}_{-70}$ & 0.10$^{+0.07}_{-0.03}$ & {2.0 $\pm$ 0.3} \\\\

\ \ 189.29834 & 62.21367 & 4.93 & 14.32$^{+0.98}_{-0.82}$ & 13.5$^{+0.5}_{-0.6}$ & 30.2 $\pm$ 0.2 & 15 & 0.9 & 7.93$^{+0.23}_{-0.45}$ & 1.3$^{+0.3}_{-0.3}$ & 1.25$^{+0.3}_{-0.22}$ & 136$^{+120}_{-100}$ & 0.06$^{+0.08}_{-0.04}$ & {2.0 $\pm$ 0.2
} \\\\

\ \ 189.10753 & 62.24305 & 3.42 & 13.46$^{+0.71}_{-0.73}$ & 13.01$^{+0.65}_{-0.52}$ & 29.4 $\pm$ 0.1 & 13 & 0.9 & 8.62$^{+0.12}_{-0.20}$ & 4.2$^{+1.3}_{-0.7}$ & 4.33$^{+1.58}_{-0.99}$ & 190$^{+90}_{-80}$ & 0.23$^{+0.09}_{-0.10}$ & {2.7 $\pm$ 0.3} \\\\

\ \ 189.09621 & 62.24877 & 2.93 & 11.42$^{+0.68}_{-0.42}$ & 11.5$^{+0.2}_{-0.2}$ & 29.6 $\pm$ 0.2 & 26 & 0.9 & 8.75$^{+0.17}_{-0.25}$ & 5.1$^{+2.2}_{-1.2}$ & 4.86$^{+1.81}_{-1.18}$ & 240$^{+100}_{-140}$ & 0.40$^{+0.15}_{-0.13}$ & {2.6 $\pm$ 0.4} \\\\

\hline

\end{tabular}
}
\end{table*}

Furthermore, the inferred ages for all
galaxy candidates within our sample
indicate younger stellar populations,
spanning a range of approximately
80 to 240 million years, similar to the ages found in earlier photometric detections of galaxies at $z \sim 10$ \citep{2021MNRAS.505.3336L, 2023ApJ...955...13B}.
Regarding dust extinction, the best-fit values of $A_V$ range from 0.02 to 0.45. These values are consistent with those found in $z=9-16$ galaxies in the NGDEEP survey \citep{2024ApJ...964L..24M}, which were also fitted using \texttt{BAGPIPES} and exhibited $A_V$ values between 0.04 and 1.03.
These findings collectively contribute
to a deeper understanding of the 
stellar populations and dust
properties within the observed 
galaxy sample.

\begin{figure*}
    \centering
    \includegraphics[width=1.0\textwidth]{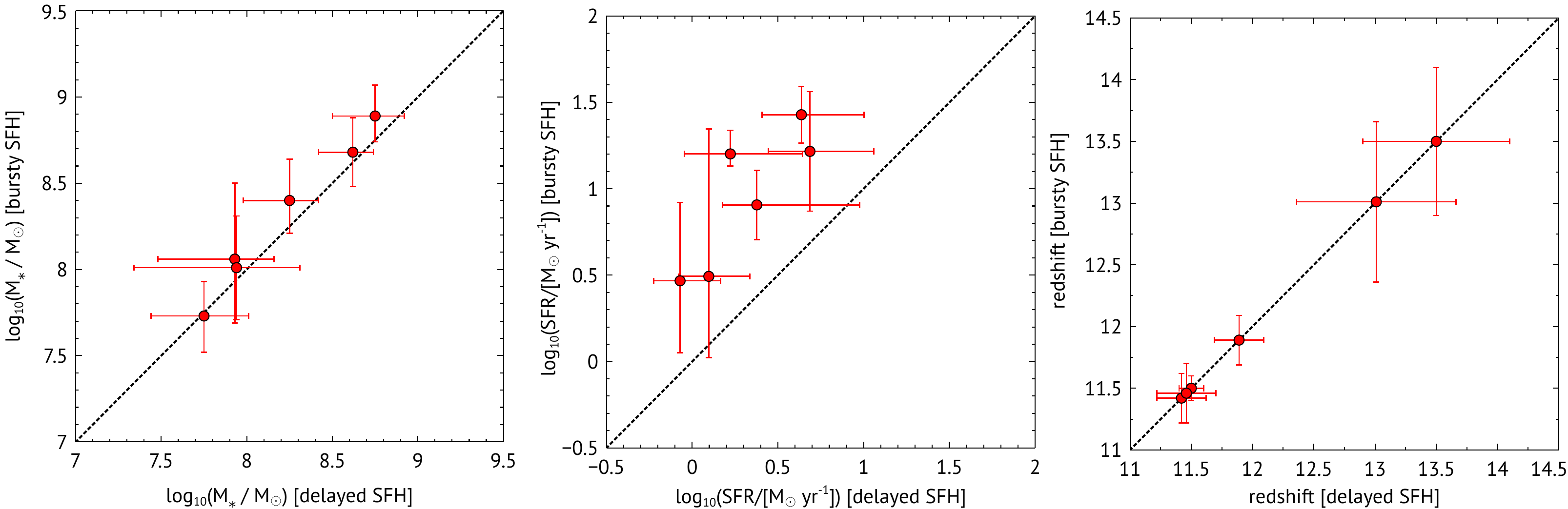}
\caption{ 
Comparison of
derived stellar masses,
star-formation rates, and
redshifts for the newly identified galaxy candidates using 
delayed-$\tau$ and bursty 
star formation history models.
Star formation rates are averaged
over the last 100 Myr.}
\label{fig:bursty_delayed_comp}
\end{figure*}

{ Recent studies 
also points to bursty star formation histories in galaxies
at $z>6$
compared to their lower-redshift counterparts, with episodic bursts temporarily enhancing the luminosity of low-mass galaxies due to the rapid formation of massive O stars 
\citep{2025A&A...697A..88L}. 
At higher redshifts
($z>10$), 
several theoretical models have demonstrated that an evolving
burstiness can naturally explain the unexpectedly high abundance of bright galaxies detected by JWST in the 
$z=10-14$ range 
\citep{2023MNRAS.522.3986F}. 
These models suggest that short, 
intense episodes of star formation—dominate the early buildup of stellar mass in the first few hundred million years after the Big Bang.
Therefore, 
to further assess the impact of star formation history assumptions, 
we re-derived the stellar mass, redshift, and SFR of our galaxy candidates using a bursty SFH model implemented in {\tt BAGPIPES}. 
We implemented the same fitting method as discussed earlier for the delayed-$\tau$ SFH model. 
Figure \ref{fig:bursty_delayed_comp} presents a comparison between the delayed-$\tau$ and bursty SFH results. 
The derived stellar masses and redshifts are consistent across both models. 
However, in 4 out of 6 galaxy candidates, the SFRs inferred from the bursty SFH are significantly higher than those obtained with the delayed-$\tau$ model.
}

\subsection{SFR vs. stellar mass relation}

\citet{2007ApJ...660L..43N} and \citet{2007A&A...468...33E} demonstrated
that galaxies at $z \sim 1$ exhibit a 
tight correlation between 
star formation rate and stellar mass,
with a scatter of 0.2 dex and a 
rough proportionality characterized
by a logarithmic slope of 0.9. 
This proportionality is similarly
observed at $z \sim 0$ using data 
from the Sloan Digital Sky Survey 
\citep{2007A&A...468...33E}.
However, the normalization at $z \sim 0$ 
is lower, reflecting the overall
decline in cosmic star formation 
rate density over time.
This tight correlation between SFR and 
stellar mass, 
known as the main sequence (MS) of
star formation, has been 
extensively documented in the
literature 
\citep[e.g.,][]{2007ApJ...670..156D,2007ApJ...660L..43N,2007A&A...468...33E}. 
This relationship has been observed 
in high redshifts, with its
evolution parametrized up to $z \sim 6$ 
(e.g., \citealt{2014ApJS..214...15S}).
However, establishing the MS at
high redshifts poses significant 
challenges due to various systematic 
uncertainties and selection effects
inherent in compiling representative 
galaxy samples 
\citep[e.g.,][]{2015A&A...575A..96G,2020ARA&A..58..661F,2021MNRAS.501.1568F,2023MNRAS.519.3064F}. 
In order to probe the poorly understood Universe at $z \gtrsim 10$, which is beginning to unfold with
the advent of the JWST, we 
investigate the SFR-stellar mass 
(SFR--$M_\ast$)
relation within our sample and 
compare it with previous studies.

\begin{figure*}
    \centering
    
\includegraphics[width=1.02\textwidth]{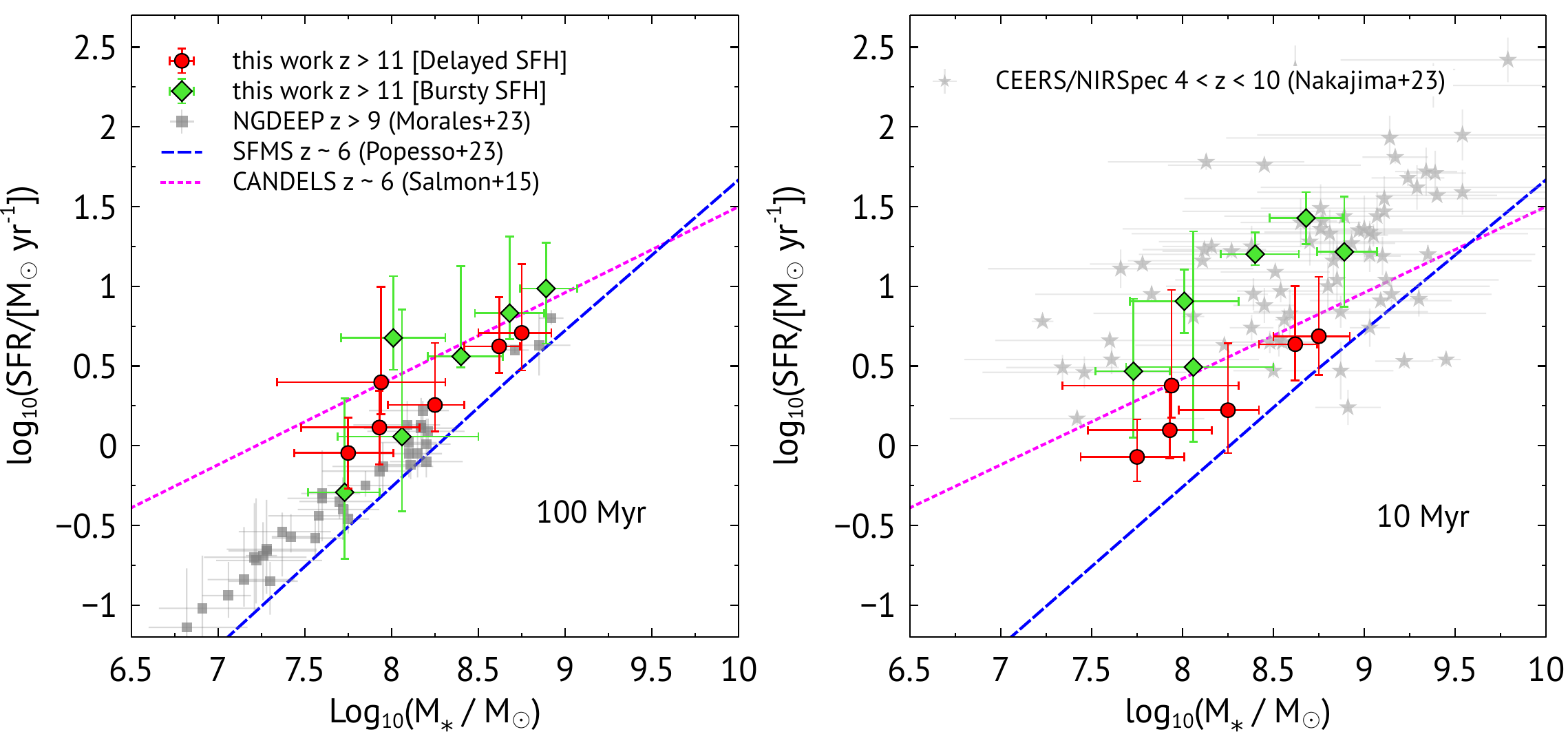}
\caption{
 Distribution of stellar mass versus star formation rate (SFR) for the photometric galaxy sample analyzed in this study.
Left: SFRs averaged over the past 100 Myr for newly identified galaxy candidates, derived using delayed (red) and bursty (green) star formation history (SFH) models. Grey squares represent NGDEEP galaxies from \citet{2024ApJ...964L..24M}, with SFRs also derived from photometry.
Right: SFRs averaged over the past 10 Myr for the same galaxy candidates, using the same SFH models (delayed in red, bursty in green). 
Grey stars show CEERS galaxies with SFRs inferred from H$\beta$ luminosity, compiled by \citet{2023ApJS..269...33N}.
In both panels, the star-forming main sequence at $z \sim 6$ is shown for reference: the blue dashed line represents the relation from \citet{2023MNRAS.519.1526P}, and the magenta dotted line from \citet{2015ApJ...799..183S}.}
\label{fig:m_star}
\end{figure*}

{ In Figure \ref{fig:m_star}, we present 
the distribution of our galaxy sample
in the SFR--$M_\ast$ plane, 
where the physical properties are derived using the {\tt BAGPIPES}
code.
For each galaxy, we compute the SFR averaged over both the most recent 100 Myr and 10 Myr time intervals, allowing us to assess recent star formation activity on different timescales.
In the left panel of Figure \ref{fig:m_star}, we compare our 100 Myr-averaged SFR--$M_\ast$ distribution to that of galaxies identified in the NGDEEP survey spanning redshifts
$z = 9-16$
\citep{2024ApJ...964L..24M}.
Their analysis employs a methodology comparable to ours, using SED fitting to estimate physical properties from deep photometric data, including 
{\tt BAGPIPES} code and 
100 Myr-averaged SFR.
This comparison provides a useful benchmark against another photometrically selected high-redshift sample.
Observationally, it has been shown that SFRs derived from 
dust attenuation corrected
H$\beta$ flux are strongly correlated with SFRs averaged over the most recent $\lesssim$10 Myr \citep{2016ApJ...833...37G,2023ApJS..269...33N}.
Motivated by this, we compare our 10 Myr-averaged SFR--$M_\ast$ relation to that of galaxies from the CEERS survey at
$z = 3-10$,
for which SFRs were measured using NIRSpec spectroscopy \citep{2023ApJS..269...33N}.
These spectroscopically confirmed galaxies provide an independent check on our photometric estimates of 
SFR--$M_\ast$ distribution over a broad redshift range.
In both comparisons, we overlay the star-forming main sequence at $z \sim 6$ 
as reported by \citet{2015ApJ...799..183S} and \citet{2023MNRAS.519.1526P}, shown as reference lines in each panel. Overall, our results are in good agreement with previous studies, supporting the robustness of our derived stellar masses and SFRs across different SFH assumptions and timescales.
}

\subsection{Limitations of SED-fitting 
method and contaminations}
We solely utilize SED fitting
tool \texttt{BAGPIPES} to derive the physical
properties of our high-redshift sample
of galaxies. 
Though the effectiveness and flexibility of
\texttt{BAGPIPES} code has been shown
in many previous studies, such as
\citet{2023ApJ...955...13B,2024ApJ...964L..24M}, the physical properties
of high-redshift galaxies derived from
\texttt{BAGPIPES} may be affected by several 
limitations and contaminations.

Analyzing 
SEDs through fitting broadband photometry, 
which mainly captures the rest-frame 
UV emissions of potential high-redshift
galaxies is one of the limitations
of our analysis. 
This wavelength range 
primarily highlights massive, 
short-lived stars within a galaxy,
which might not represent the 
majority of its stellar mass. 
Due to the limited availability of
rest-frame optical photometry, 
SED fitting is expected to 
underestimate stellar masses. 
Specifically,  
\citet{2021MNRAS.501.1568F} demonstrated 
that relying solely on UV photometry 
can result in an underestimation of 
stellar masses by up to 0.6 dex. 
Stellar masses can be accurately derived
from the UV continuum only if a galaxy 
is $\lesssim$ 10 Myr old,
which is the typical lifetime 
of massive stars. 
Beyond this age, accurate stellar
mass determination requires continuum 
detection redward of the Balmer break. 
Furthermore, relying solely on rest-frame 
UV photometry can create ambiguities
between stellar mass, SFR, and age, 
as noted by \citet{2023MNRAS.519.3064F}.
{ To confirm if galaxies at
$z \gtrsim 10$ possess an evolved 
stellar population, as suggested by
recent simulations \citep{2023MNRAS.521..497M}, 
JWST/NIRCam imaging at wavelengths 
$\lambda \leq 5 \mu$m should be 
supplemented with future
observations targeting
the rest-frame optical 
emission of these very 
high-redshift galaxies.}

Another constraint of our analysis 
arises from assuming a Lyman continuum
(LyC) escape fraction of
$f_{\rm esc} = 0$ as per \texttt{BAGPIPES}. 
However, the LyC escape fraction 
significantly influences the shape
of the SED through its impact 
on nebular emission. 
Specifically, it attenuates both nebular
line and continuum emissions,
resulting in a bluer SED and a 
steeper UV slope 
\citep{2019MNRAS.490..978P}. 
Observations of known LyC leakers 
at low redshifts reveal a clear 
correlation between UV slope and 
LyC escape fraction 
\citep{2022MNRAS.517.5104C}. 
Although recent studies suggest
low $f_{\rm esc}$(LyC) values of
$\lesssim 10\%$ in galaxies at $z > 7$ 
\citep{2023ApJ...944...61W}, 
it remains essential to account 
for the effect of $f_{\rm esc}$(LyC) 
in highly star-forming primeval galaxies
in the early Universe.
This consideration is paramount as
it could pose challenges in 
replicating the extremely blue 
UV slopes observed in some galaxies
detected by JWST at $z > 10$. 
Therefore, future investigations 
aimed at accurately determining 
physical parameters of high-redshift 
galaxies through SED-fitting must 
incorporate templates that allow
for $f_{\rm esc} > 0$.

{
The primary source of contamination in 
broad-band photometry of high-redshift
galaxy samples often stems from 
faint low-redshift galaxies with strong line emission or Balmer breaks. Although our selection criteria 
for high-redshift samples are 
meticulously crafted to
mitigate contamination from
lower-redshift interlopers,
the potential still exists for 
some of our candidates to reside
at lower redshifts. 
Spectroscopic confirmation is
imperative to accurately determine
the redshifts of these candidates.
}

\section{Conclusion}\label{con}

In this paper, we present the outcomes of our investigation aimed at identifying $z>11$ galaxy candidates  in the JWST NIRCam observations of JADES GOODS-S and GOODS-N fields. We undertake an extensive data reduction process that involves several custom procedures. 
We use two distinct codes, \texttt{EAZY} and \texttt{BAGPIPES}, for photometric redshift determination and investigating the properties of these galaxies.
 We detect six $z>11$ candidate galaxies distributed as follows: three candidates between $11 < z < 12$, and four candidates between 
 $12 < z < 15$. 
All candidates are robust high-redshift detections with 70-100\% of their probability within $\Delta z = 1$, meeting the classification `high-quality' used in a recent \textit{JWST} UNCOVER high-redshift survey \citep{2023MNRAS.524.5486A}.
The detection of these young galaxies spans a diverse range of magnitudes, with 
F277W AB Kron magnitude values between 28.5 and 30.4, consistent with previous JADES surveys, which achieved detections as faint as AB mag 31.0 for the young galaxies at redshifts greater than 10 \citep{2024ApJ...964...71H}. 

To assess the physical properties of the galaxies, we perform SED fitting to the photometric data of these galaxies using \texttt{BAGPIPES}.
We obtain measurements for stellar mass, SFRs averaged over the last 100 Myr
and 10 Myr, 
dust content, and mass-weighted age, which are listed in 
Table \ref{t:bagpipes}.
We find the intrinsic stellar masses of our galaxy candidates to vary within the range of $\log (M_*/M_{\odot}) = 7.75$ to $8.75$. Their SFRs span between 0.9 and 5.1 $M_{\odot}\text{yr}^{-1}$, while their sSFRs, quantified as $\log \text{sSFR}/\text{Gyr}$, range from 0.95 to 1.46. 
The mass-weighted ages of these galaxies range between 80 to 240 Myr, with their dust content varying between $A_V$ = 0.02 and 0.45.
We present the distribution of galaxy candidates on the stellar mass versus star formation rate plane and compare our findings, which closely align with those of previous surveys \citep{2023ApJS..269...33N, 2024ApJ...964L..24M, 2024A&A...684A..75C}.

Previous studies have demonstrated that photometry is a reliable method for determining galaxy properties. When used in conjunction with SED fitting codes such as \texttt{BAGPIPES}, photometry can reliably estimate photometric redshifts, stellar masses, and star formation rates \citep{2024MNRAS.529.4728D}. However, slight differences that may arise between properties derived from spectroscopy and those obtained through photometry must be considered when calibrating an absolute scale for the star formation rates and stellar mass histories of galaxies. Moving forward, JWST/NIRSpec observations will be essential for confirming these early galaxies, allowing for a more detailed analysis of their intrinsic properties.

\section*{Acknowledgements}
We thank the anonymous referee for
their helpful comments and suggestions.
This work is based on observations made
with the NASA/ESA/CSA James Webb Space Telescope.
We thank Dr. Daniel Eisenstein for the helpful discussion on JWST/JADES data. We acknowledge support by
NASA Chandra grant GO5-26101X 16619404 and NASA ADAP Grant 80NSSC21K0637.

\section*{DATA AVAILABILITY}

The data underlying this article will be shared on reasonable request
to the corresponding author.

\bibliographystyle{mnras}
\bibliography{example} 





\bsp	
\label{lastpage}
\end{document}